\documentclass[pdflatex,amsmath,amsfonts,prd,showpacs,twocolumn,nofootinbib,superscriptaddress]{revtex4-2}

\usepackage{graphicx}
\usepackage{color}
\usepackage{physics,bm}
\usepackage{ulem}

\def\beq{\begin{equation}}
\def\eeq{\end{equation}}
\def\bea{\begin{eqnarray}}
\def\eea{\end{eqnarray}}
\def\barr{\begin{array}}
\def\earr{\end{array}}

\newcommand{\bvec}[1]{\boldsymbol{#1}}

\begin{document}

\title{Lattice study of SU(2) gauge theory coupled to four adjoint Higgs fields} 

\author{Guilherme Catumba}
\affiliation{Instituto de Física Corpuscular (IFIC), CSIC-Universitat de Valencia, 46071, Valencia, Spain}
\author{Atsuki Hiraguchi}
\affiliation{Institute of Physics, National Yang Ming Chiao Tung University,
1001 Ta-Hsueh Road, Hsinchu 30010, Taiwan}
\affiliation{CCSE, Japan Atomic Energy Agency, 178-4-4, Wakashiba, Kashiwa, Chiba 277-0871, Japan}
\author{{Wei-Shu} Hou}
\affiliation{Department of Physics, National Taiwan University, Taipei 10617, Taiwan}
\author{Karl Jansen}
\affiliation{Deutsches Elektronen-Synchrotron DESY, Platanenallee 6, 15738 Zeuthen, Germany}
\author{Ying-Jer Kao}
\affiliation{Department of Physics, National Taiwan University, Taipei 10617, Taiwan}
\affiliation{Center for Theoretical Physics and Center for Quantum Science and Technology, National Taiwan University, Taipei, 10607, Taiwan}
\author{C.-J. David Lin}
\affiliation{Institute of Physics, National Yang Ming Chiao Tung University, 1001 Ta-Hsueh Road, Hsinchu 30010, Taiwan}
\affiliation{Center for High Energy Physics, Chung-Yuan Christian University, Chung-Li 32023, Taiwan}
\author{Alberto Ramos}
\affiliation{Instituto de Física Corpuscular (IFIC), CSIC-Universitat de Valencia, 46071, Valencia, Spain}
\author{Mugdha Sarkar}
\affiliation{Department of Physics, National Taiwan University, Taipei 10617, Taiwan}
\affiliation{Physics Division, National Centre for Theoretical Sciences, Taipei 10617, Taiwan}

\begin{abstract}
    Gauge theories with matter fields in various representations play an important role in different branches of physics. Recently, it was proposed that several aspects of the interesting pseudogap phase of cuprate superconductors near optimal doping may be explained by an emergent $SU(2)$ gauge symmetry. Around the transition with positive hole-doping, one can construct a $(2+1)-$dimensional $SU(2)$ gauge theory coupled to four adjoint scalar fields which gives rise to a rich phase diagram with a myriad of phases having different broken symmetries. We study the phase diagram of this model on the Euclidean lattice using the Hybrid Monte Carlo algorithm. We find the existence of multiple broken phases as predicted by previous mean field studies. Depending on the quartic couplings, the $SU(2)$ gauge symmetry is broken down either to $U(1)$ or $\mathbb{Z}_2$ in the perturbative description of the model. We further study the confinement-deconfinement transition in this theory, and find that both the broken phases are deconfining in the range of volumes that we studied. However, there exists a marked difference in the behavior of the Polyakov loop between the two phases.
\end{abstract}

\maketitle

\section{Introduction}
\label{sec:intro}
Understanding the interplay between local and global symmetries have been a topic of interest in theoretical physics. In this regard, gauge-Higgs theories play an important role in both particle  and condensed matter physics. The study of such theories in different dimensions with different gauge groups and with Higgs fields in various representations of the gauge group, provide a better understanding of several fundamental aspects of  physics. For instance, gauge-Higgs theories describe the electroweak sector of the Standard Model of particle physics \cite{Salam:1959zz,Glashow:1961tr, Weinberg:1967tq} and form the basis of the BCS theory of superconductivity in condensed matter physics \cite{Bardeen:1957kj,Bardeen:1957mv}.

Since its discovery \cite{Wilson:1974sk}, lattice gauge theory techniques have provided a controlled systematic method for non-perturbative investigation of gauge theories coupled to matter fields. Due to their accessibility in the early days of lattice computations, non-Abelian gauge theories in $d=3$ dimensions were studied in great detail as effective theories for $d=4$ field theories at high temperature. In particular, the $d=3$ $SU(2)$ gauge theory with an adjoint Higgs field is obtained by a dimensional reduction of $d=4$ pure $SU(2)$ Yang-Mills theory in the deconfined phase at high temperature \cite{Appelquist:1981vg}. It was predicted that this theory would have a spontaneous gauge symmetry breaking pattern\footnote{Although this is the conventional way of writing in literature, the gauge symmetry can never be spontaneously broken \cite{Elitzur:1975im}. It should be assumed that the gauge has been appropriately fixed. We will use the conventional notation in this work for convenience. We also use the terms ``broken phase" and ``Higgs phase" interchangeably.} $SU(2)\to U(1)$ and the symmetric and broken (Higgs) phases would be separated completely by a phase boundary, unlike the case of Higgs fields in the fundamental representation where the two phases are connected  \cite{Fradkin:1978dv}. Subsequent numerical studies of the $d=3$ $SU(2)$ gauge theory with a single adjoint Higgs field \cite{Nadkarni:1989na,Hart:1996ac,Kajantie:1997tt, Hart:1999dj} revealed a phase diagram consisting of a symmetric phase and a Higgs phase which are partially separated by a line of first-order transitions. This raised the question whether both the phases are confining.

The $d=3$ $SU(2)$ gauge theory with one adjoint Higgs, also known as the $3d$ Georgi-Glashow model, is well-known for the existence of topologically non-trivial solutions of the classical field equations, the 't Hooft-Polyakov monopole \cite{tHooft:1974kcl,Polyakov:1974ek}. It was shown through semi-classical methods that the Higgs phase is confining due to the condensation of monopoles \cite{Polyakov:1976fu,Mandelstam:1974pi}. Nonperturbative lattice studies of the monopole density and mass have supported this picture that the Higgs phase and symmetric phase are both confining and continuously connected in the phase diagram \cite{Davis:2001mg,Niemi:2022bjg}.

Introducing more adjoint Higgs fields lead to different global and local symmetry breaking realizations. For the number of adjoint Higgs fields, $N_h>3$ (the number of generators of the $SU(2)$ gauge group), {a further gauge symmetry breaking pattern $SU(2)\to \mathbb{Z}_2$ can occur} \cite{Fradkin:1978dv, Bonati:2021oqq}. The $N_h=4$ adjoint Higgs theory was first studied using numerical simulations in the strong gauge coupling limit where a global $O(4)$ symmetric potential was considered \cite{Scammell:2019erm}. A subsequent numerical study with dynamical gauge fields considered the $N_h=4$ case again with a global $O(4)$ symmetry \cite{Bonati:2021tvg}. Both studies provide evidence for two broken Higgs phases with different broken symmetries.

Recent interests in the emerging phenomena in quantum matter calls for a better understanding of the role of gauge symmetries in effective models. In particular, a plethora of exotic phenomenon emerges from the experiments on the hole-doped cuprates around the dome of high temperature superconductivity~\cite{Proust:2019eu}. Scanning tunneling microscopy  observations have revealed the presence of nematic and charge density orders that vanish near optimal doping. Also, it is observed that the electron spectral functions transform from “Fermi arcs” to a full Fermi surface across optimal doping~\cite{2014.Fujita,2014.He}. The $d=3$ $SU(2)$ gauge theory with $N_h=4$ adjoint Higgses has been proposed as an effective theory for the pseudogap phase to explain these observations \cite{Sachdev:2018nbk}. 

In this study, we extend the investigations of the $SU(2)$ lattice gauge theory with $N_h=4$ adjoint Higgs fields with the complete Higgs potential proposed in \cite{Sachdev:2018nbk}. We find that the system hosts a rich variety of broken Higgs phases with two gauge symmetry breaking patterns, $SU(2)\to U(1)$ and $SU(2)\to \mathbb{Z}_2$, consistent with the mean field predictions in \cite{Sachdev:2018nbk}. Furthermore, we calculate the Polyakov loop, which is an order parameter for deconfinement of particles having a fundamental gauge charge, in different phases. We find interesting behavior for the Polyakov loop in the two differently broken Higgs phases. We observe that there are broken phases which deconfine and can be phenomenological candidates for the pseudogap region of cuprate superconductors {where spinon and chargon degrees of freedom may decouple.}

This paper is organized as follows. In Sec.~\ref{sec:adj_higgs}, we discuss the action of the $SU(2)$ gauge theory with 4 adjoint Higgs fields and properties of the phase diagram known from mean-field studies. We describe the observables used in our study in Sec.~\ref{sec:observables}. In Sec.~\ref{sec:results}, we discuss our numerical results of the phase diagram and finally conclude in Sec.~\ref{sec:conclusion}. In two appendices, we present further details on the global symmetries of the Higgs potential and forces computed for the Hybrid Monte Carlo simulation. 
\section{SU(2) gauge theory with four adjoint Higgs}
\label{sec:adj_higgs}
Following the notation of \cite{Hart:1996ac}, the lattice action of the $d=3$ $SU(2)$ gauge theory coupled to $N_h=4$ adjoint Higgs fields is given as
\begin{align}
    S &= \beta \sum_x \sum_{\mu < \nu}^3 \left(1 - \frac{1}{2} \Tr U_{\mu\nu}(x) \right) \nonumber\\ 
    &- 2\kappa \sum_{x,\mu} \sum_{n=1}^{N_h} \Tr \left( \Phi_n(x) U_\mu(x)\Phi_n(x+\hat\mu) U^\dagger_\mu(x)\right) \nonumber \\
    &+ \sum_x V(\{\Phi_n(x)\})\,, \label{Eq:lattaction}
\end{align}
where $U_\mu(x)$ is the lattice $SU(2)$ gauge link, $U_{\mu\nu}$ denotes the plaquette in the $\mu$--$\nu$ plane and the adjoint Higgs fields are written as $2\times 2$ matrices, $\Phi_n = \Phi_n^\alpha {\sigma^\alpha/2}$ ($n=1,\ldots ,N_h$), $\sigma^\alpha$ being the Pauli matrices. 
With the shorthand notation,
\begin{equation}\label{Eq:Bmn}
    B_{mn}(x)=\sum_{\alpha=1}^3 \Phi_m^\alpha(x)\Phi_n^\alpha(x),
\end{equation}
the Higgs potential is given as,
\begin{align}
    &V(\{\Phi_n\}) \nonumber \\
    &= \sum_{n=1}^{N_h} B_{nn} + \lambda \Big( \sum_{n=1}^{N_h} \left( B_{nn} - 1\right)^2 + \sum_{n \neq m}^{N_h} B_{nn}B_{mm} \Big) \nonumber \\
  &+ \frac{\hat{u_1}}{2}  \Bigg[ \frac{1}{2}\sum_{n=1}^{N_h} (B_{nn})^2 
  + B_{11}B_{22} + B_{33}B_{44} \nonumber \\
  &~~~~~- B_{11}B_{33} - B_{11}B_{44} - B_{22}B_{33} - B_{22}B_{44} \Bigg] \nonumber\\
  &+ \hat{u_2} \Bigg[\frac{1}{2}\sum_{n=1}^{N_h} (B_{nn})^2 
  + 2(B_{12})^2 + 2(B_{34})^2 - B_{11}B_{22} - B_{33}B_{44} \Bigg] \nonumber\\
  &+ \hat{u_3} \sum_{n \neq m}^{N_h} B_{nm}B_{nm} \,, \label{Eq:lattpot}
\end{align}
where the spacetime dependence has been suppressed. 

\begin{figure*}[t]
  \centering
   \includegraphics[width=0.4\textwidth]{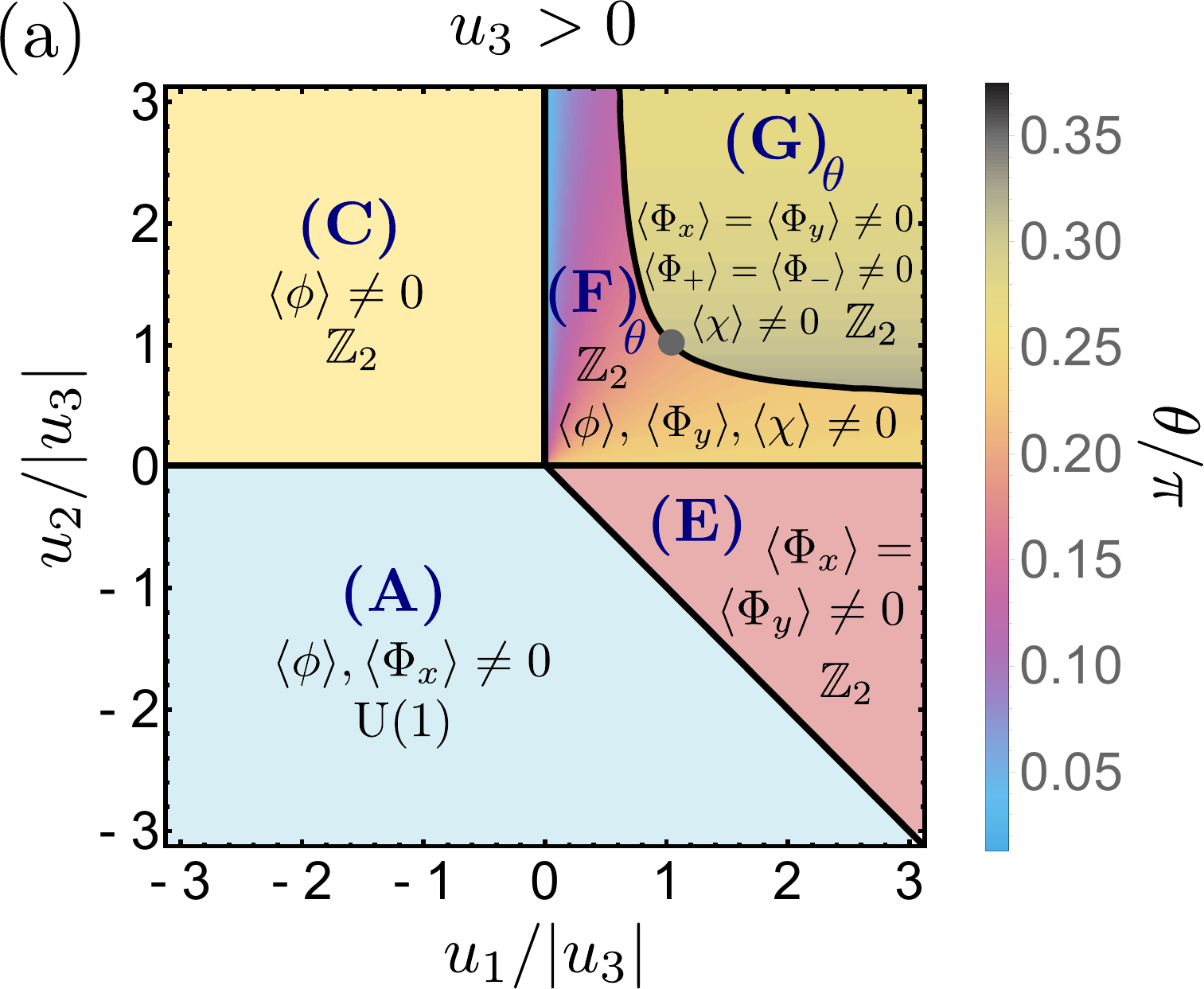}
   \includegraphics[width=0.3\textwidth]{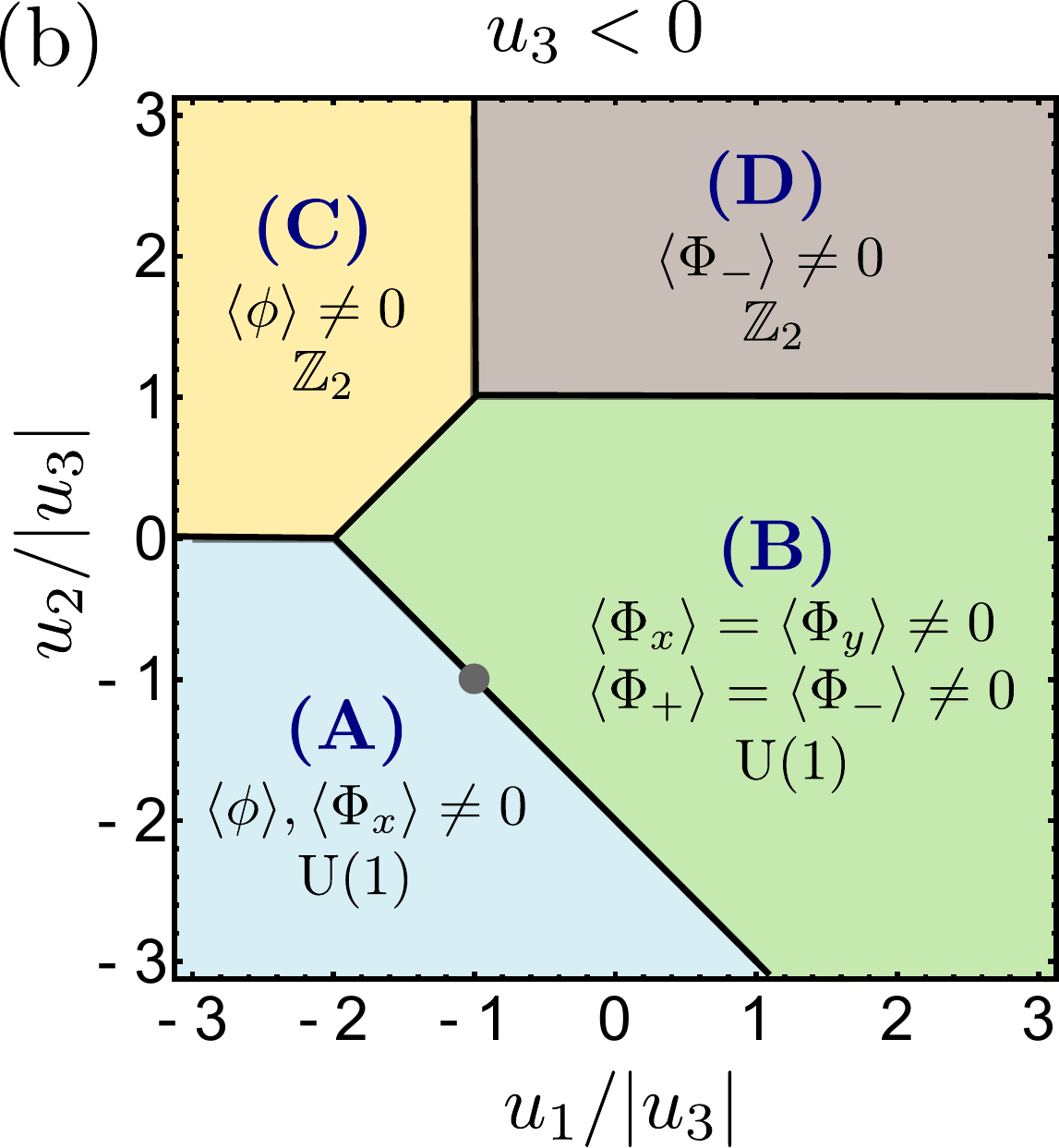}
	\caption{Mean-field prediction of the phase diagram as function of the quartic coupling ratios, $u_1/|u_3|$ and $u_2/|u_3|$, for, a) $u_3 > 0$ and b) $u_3<0$. The continuum couplings $s$ and $u_0$ have been chosen appropriately to remain in the broken symmetric region. The broken phases labelled (A), (B), $\ldots$, (G) are shown alongwith the behavior of the bilinear order parameters and the remnant local symmetry obtained in a perturbative treatment. {The figures have been reproduced from Ref.~\cite{Sachdev:2018nbk}.}
	} \label{fig:meanfld}
\end{figure*}

To explain the connection between the lattice notation above and the notation in \cite{Sachdev:2018nbk}, we reproduce the continuum Lagrangian,
\begin{align}
  \mathcal{L}_{\mathcal{H}} = &\frac{1}{4 g^2} \bvec{F}_{\mu\nu} \cdot
 \bvec{F}_{\mu\nu} 
 + \sum_{i=x,y} \left| \partial_\mu \bvec{\mathcal{H}}_i - \bvec{A}_\mu \times \bvec{\mathcal{H}}_i \right|^2 \nonumber \\
 &+ V(\bvec{\mathcal{H}}_{x,y};s,u_i(i=0,\ldots,3))\, , \label{Eq:cont_action}
\end{align}
where the field strength $\bvec{F}_{\mu\nu}$ is given in terms of the $SU(2)$ gauge field $\bvec{A}_\mu$ as $\bvec{F}_{\mu\nu}=\partial_\mu \bvec{A}_\nu - \partial_\nu \bvec{A}_\mu - \bvec{A}_\mu \times \bvec{A}_\nu$.
We make explicit the components of the complex Higgs fields $\bvec{\mathcal{H}}_{x/y}$ as $\bvec{\mathcal{H}}_x = \varphi_1 + i \varphi_2$ and $\bvec{\mathcal{H}}_y = \varphi_3 + i \varphi_4$. To construct the lattice action, we define dimensionless Higgs fields $\Phi_n^\alpha (x) = \sqrt{a/\kappa} \varphi_n^\alpha$ on each site $x$ of a Euclidean 3-dimensional lattice with lattice spacing $a$, where $n=1${--}4 denotes the Higgs flavor index and $\alpha=1${--}3 is the $SU(2)$ color index. The lattice Higgs fields have been scaled by the dimensionless coupling $\kappa$, also known as the hopping parameter. The continuum $SU(2)$ algebra-valued gauge fields, $\bvec{A}_\mu(x)$, are represented as $SU(2)$ group-valued gauge links, $U_{\mu}(x) = \mathrm{exp} (iag\bvec{A}_\mu (x))$, residing on the link connecting neighboring lattice sites $x$ and $x+\hat\mu$, where $\hat\mu$ denotes the unit vector along the 3 spacetime directions.    
The dimensionless lattice couplings are related to the couplings in the continuum Lagrangian in Eq.~(2.1) of \cite{Sachdev:2018nbk} as
\begin{gather}
   \beta = 4/ag^2, \qquad \lambda=u_0 a \kappa^2, \nonumber \\
  s = \frac{1}{a^2}\left[\frac{1}{\kappa} - 3 - 
  \frac{2\lambda}{\kappa}\right], \quad
  \hat{u_i} = u_i a \kappa^2 \quad (i=1,2,3). \label{Eq:lattcouplings}
\end{gather}

The terms in the potential above have been chosen to satisfy the time-reversal and square-lattice symmetries of the system \cite{Zhang:2002zz,Sachdev:2018nbk}. The above symmetries are mapped to continuous or discrete global symmetries of the gauge-Higgs action in Eq.~(\ref{Eq:lattaction}). We discuss these symmetries in more detail in Appendix \ref{app:symm}. The spontaneous breaking of various combinations of these global symmetries of the potential result in a rich phase diagram which has been studied in the mean-field limit \cite{Sachdev:2018nbk,DePrato:2006jx}. For $u_1=u_2=u_3$, the potential becomes invariant under a global $O(4)$ symmetry and this special case was studied in \cite{Scammell:2019erm, Bonati:2021oqq}. 

We briefly explain here the physics why this $SU(2)$ theory with $N_h=4$ adjoint Higgs serves as an effective theory of the cuprate system \cite{Sachdev:2018nbk}.
Assuming the excitations near the Fermi surfaces are unfractionalized electrons, it is possible to write down an $SU(2)$ gauge theory of  fluctuating incommensurate  spin density wave (SDW) fluctuations.   
By placing the SDW order parameter in a local rotating reference frame, it results in a $SU(2)$ gauge field coupled to four Higgs fields transforming in the adjoint representation of the gauge group. The charge degrees of freedom are further coupled to the Higgs fields. 
In a perturbative treatment with gauge-fixing, the $SU(2)$ gauge symmetry is spontaneously broken due to the formation of the Higgs vev and this induces the formation of charge ordering in the cuprate system. 
The symmetric phase corresponds to the metallic Fermi liquid phase of cuprate systems while the broken phases correspond to the pseudogap phase which shows different spin and charge order.

A mean-field analysis of the theory found six different broken phases in the parameter space spanned by the scalar quartic couplings in Eq.~(\ref{Eq:lattpot}). These phases are labelled by (A), (B),$\ldots$, (G) in Ref.~\cite{Sachdev:2018nbk}. Phases (A) and (B) have remnant $U(1)$ symmetry while phases (C)-(G) have remnant $\mathbb{Z}_2$ symmetry \cite{DePrato:2006jx,Sachdev:2018nbk}. To make the discussion self-contained, we reproduce the mean-field phase diagram of the Higgs phases concisely plotted in terms of the ratios of the quartic couplings, $u_1/|u_3|$, $u_2/|u_3|$ and the sign of $u_3$ in Fig.~\ref{fig:meanfld}, taken from Ref.~\cite{Sachdev:2018nbk}.
The broken phases can be characterized by the non-zero expectation values of local gauge-invariant Higgs field bilinears which signal breaking of various global symmetries. However, it should be noted that it is not possible to construct local gauge-invariant order parameters that signal the gauge symmetry ``breaking" patterns {---} $SU(2)\to U(1)$, or $SU(2)\to \mathbb{Z}_2$.

The theory shows interesting confinement properties in the phase diagram. The symmetric $SU(2)$ phase is obviously confining without any explicit breaking of the $\mathbb{Z}_2$ center symmetry due to the absence of matter fields in the fundamental representation. The center symmetry is spontaneously broken in the Higgs phase leading to deconfinement. However, the phase with remnant $U(1)$ symmetry admits 't Hooft-Polyakov monopole solutions which ultimately lead to confinement in large length scales \cite{Polyakov:1976fu,Mandelstam:1974pi}.

The Higgs phase with remnant $\mathbb{Z}_2$ symmetry is of interest due to its connection to 3-dimensional $\mathbb{Z}_2$ lattice gauge theory. In the limit $\kappa\to\infty$, the action in Eq.~(\ref{Eq:lattaction}) effectively becomes the action for a $\mathbb{Z}_2$ gauge theory consisting only of the plaquette term made up of gauge links $U_{x\mu} = \lambda_{x\mu} \in \mathbb{Z}_2$ \cite{Bonati:2021tvg}. The latter theory has a transition at intermediate $\beta$ separating a confining phase where the Wilson loops constructed out of $\mathbb{Z}_2$ gauge links obeys an area law, and a deconfining phase with Wilson loops following a perimeter law \cite{Wegner:1971app}. However, a modern interpretation of this transition is given in terms of non-trivial topological order which exists only in the deconfining phase \cite{Sachdev:2018ddg}. Thus, it would be interesting to observe if these properties would persist for finite $\kappa$ in the $SU(2)$ gauge theory with adjoint Higgs.

\section{Observables for probing the phase structure}
\label{sec:observables}
We are interested in the classification of the various Higgs phases present in the phase diagram. They are related to the breaking of the various global symmetries described in Appendix \ref{app:symm}.
Due to gauge symmetry, one should consider observables which are gauge-invariant. 
At lowest order, one can construct various color-neutral bilinear combinations 
of the $N_h=4$ adjoint Higgs fields, $B_{mn}$, defined in Eq.~\ref{Eq:Bmn}.
It was shown in Ref.~\cite{Sachdev:2018nbk} that the following combinations of  
$B_{mn}$ serve as order parameters for the global flavor symmetries of the Higgs potential in Eq.~(\ref{Eq:lattpot}), 
\begin{align}
    \phi &= B_{11} + B_{22} - B_{33} - B_{44}  ,\nonumber \\
    \Phi_x &= B_{11} - B_{22} + 2iB_{12} , \nonumber \\
    \Phi_y &= B_{33} - B_{44} + 2iB_{34} , \nonumber \\
    \Phi_+ &= B_{13} - B_{24} + i(B_{14} + B_{23}), \nonumber \\
    \Phi_- &= B_{13} + B_{24} + i(B_{14} - B_{23}),
    \label{Eq:orderparams}
\end{align}
It should be noted that the observables $\Phi_{x/y}$ and $\Phi_{\pm}$ are defined as complex numbers and a non-zero value of these observables due to a broken global symmetry can equivalently mean either the real part or imaginary part or both could be non-zero. Moreover, the observable $\phi$ can be either positive or negative when the corresponding global symmetry is broken. Therefore, we plot the absolute values for these observables in our results.

In the presence of charged electron-like states coupled to the Higgs condensate, these order parameters signal different kinds of charge ordering. The order parameter $\phi$ signals {an} Ising nematic order, while $\Phi_x$ and $\Phi_y$ denote CDW ordering along incommensurate orthogonal wave vectors $2\bvec{K}_x$ and $2\bvec{K}_y$ respectively.
The bilinear $\Phi_\pm$ indicates CDW order at wave vector $\bvec{K}_x \pm \bvec{K}_y$.  

One can also define a set of local Higgs field observables \cite{scheurer_chi} that act as order parameters for the global $\mathbb{Z}_2$ symmetry ($\Phi \leftrightarrow -\Phi$) of the action as 
\begin{align}
    \chi_{xyy}&
    = -2\epsilon_{\alpha\beta\gamma}\left( \Phi_2^\alpha\Phi_3^\beta\Phi_4^\gamma - i \Phi_1^\alpha\Phi_3^\beta\Phi_4^\gamma\right)\,, \label{Eq:chixyy} \\
    \chi_{yxx}& 
    = -2\epsilon_{\alpha\beta\gamma}\left( \Phi_4^\alpha\Phi_1^\beta\Phi_2^\gamma - i \Phi_3^\alpha\Phi_1^\beta\Phi_2^\gamma\right). \label{Eq:chiyxx}
\end{align}
where $\epsilon$ is the Levi-Civita symbol.
These correspond to the breaking of the time-reversal symmetry in the cuprate system and are expected to be zero in all the phases except the (F) and (G) Higgs phases. 
It should be noted that a similar observable, $\chi$, defined in \cite{Sachdev:2018nbk},
\begin{equation}
    \chi_{ijk} = \bvec{H}_i \cdot (\bvec{H}_j \times \bvec{H}_k)\, ,
\end{equation}
where the indices $i,j,k$ indicate sites of the $d=2$ spatial lattice, cannot be used in our non-perturbative study because it is not gauge-invariant. The field $\bvec{H}_i$ at lattice site $i$ is defined as $\bvec{H}_i=\Re{\left[\bvec{\mathcal{H}}_x(i) e^{i\bvec{K}_x\cdot \bvec{r}_i} + \bvec{\mathcal{H}}_y(i) e^{i\bvec{K}_y\cdot \bvec{r}_i}\right]}$, where $\bvec{r}_i$ denote the spatial coordinates of site $i$.

The Polyakov loop, on a $N_s^2\times N_t$ lattice, is defined on each spatial site as,
 \begin{equation}
    L(\vec{x}) = \Tr \prod_{t=1}^{N_t} U_t(\vec{x},t)\,,
\end{equation} 
 where $U_t(\vec{x},t)$ is the gauge link along the temporal direction at the lattice site with coordinates $\vec{x},t$. We measure the spatial average of the Polyakov loop,
 \begin{equation}
 L=\frac{1}{N_s^2} \sum_{\vec{x}}  L(\vec{x})\,,
 \end{equation}
 as an order parameter for the confinement-deconfinement transition. 
 It describes a single static infinitely heavy test charge and its expectation value is proportional to $\mathrm{exp}(-F/T)$, where $F$ is the free energy of the test charge and $T$ is the temperature. In the confined phase, it costs infinite energy $F$ to create a single test charge and hence, $\expval{L}=0$. In the deconfined phase, there is a finite $F$ for a test charge and $\expval{L}\neq 0$ at finite temperature. In this theory, the Polyakov loop can also be understood as the order parameter for the spontaneous breaking of the global $\mathbb{Z}_2$ symmetry (center group of $SU(2)$) of the theory. In the deconfined phase with spontaneously broken $\mathbb{Z}_2$ center symmetry, the Polyakov loop can take either a positive or negative finite value. In a finite-volume on the lattice, where the number of degrees of freedom is finite, tunnelings between these two states can occur frequently. Therefore, in numerical simulations, one usually measures the absolute value, $\expval{|L|}$, in this phase.
\section{Numerical results and discussion}
\label{sec:results}
\subsection{Probing various Higgs phases}
\begin{figure}[t]
  \centering
   \includegraphics[width=0.45\textwidth]{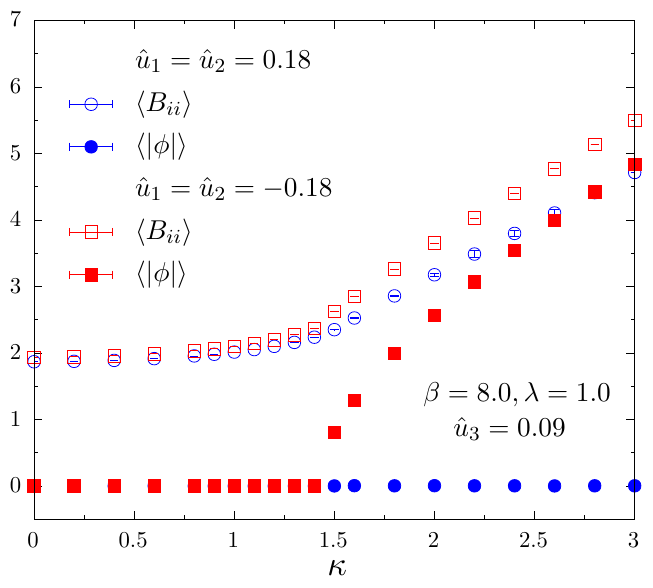}
  \caption{Variation of $\expval{B_{ii}}$ and $\expval{|\phi|}$ as a function of $\kappa$ for different values of the quartic couplings $\hat{u}_1$ and $\hat{u}_2$. The results have been obtained on $N_s=N_t=16$ lattices.}\label{fig:rhosq}
 \end{figure} 
 
We discuss results from our Monte Carlo simulation of the phase diagram of the $SU(2)$ gauge theory with $N_h=4$ adjoint Higgs fields. We have employed the Hybrid Monte Carlo (HMC) algorithm to simulate the theory and measure the gauge-invariant bilinears discussed in Sec.~\ref{sec:observables}. The majority of the results for the Higgs phase diagram have been carried out in lattices of size $N_s=N_t=12$. Further details about the HMC simulation are provided in Appendix \ref{app:hmc}. Our results qualitatively confirms the mean-field phase diagram described in Ref.~\cite{Sachdev:2018nbk}. Preliminary results from our study have appeared in \cite{Sarkar:2023lL,Catumba:2023srt}.

\begin{figure}
  \centering
   \includegraphics[width=0.33\textwidth]{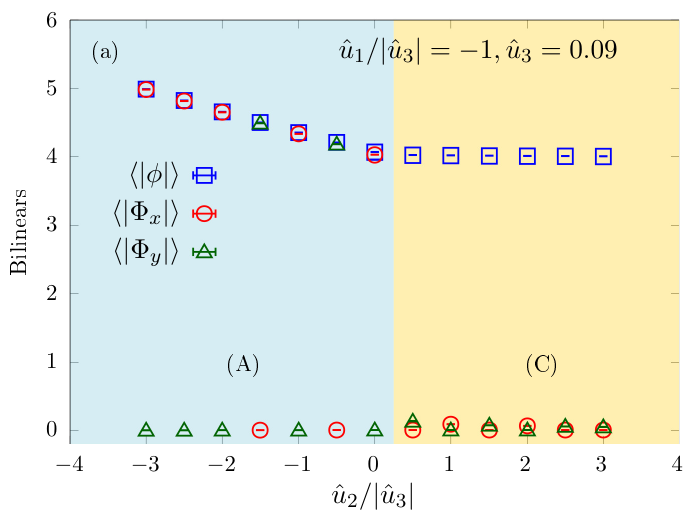}
   \includegraphics[width=0.33\textwidth]{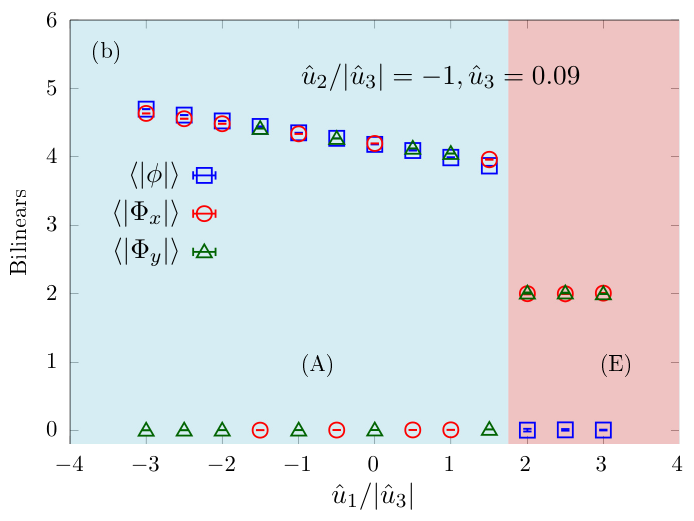}
   \includegraphics[width=0.33\textwidth]{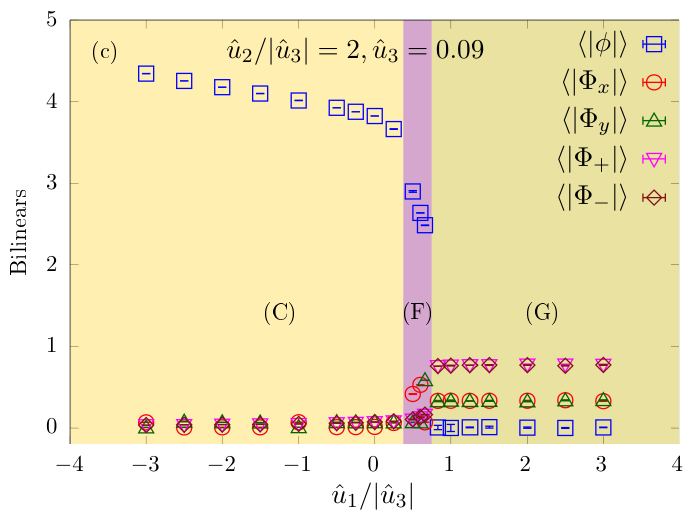}
   \includegraphics[width=0.33\textwidth]{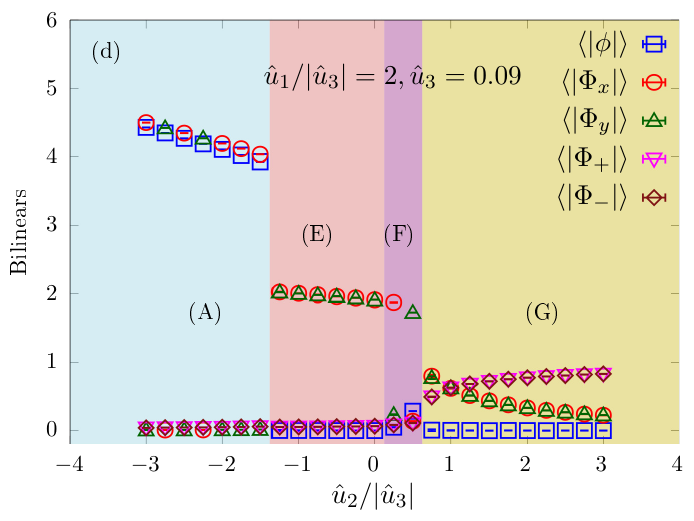}  
  \caption{Variation of the bilinears as a function of the quartic couplings, $\hat{u}_1$ or $\hat{u}_2$, at fixed $\hat{u}_3>0$. The results have been obtained on lattices of size $N_s=N_t=12$.}\label{fig:pos_u3}
 \end{figure} 

As mentioned in Sec.~\ref{sec:intro} for $d=3$ $SU(2)$ gauge theory with $N_h=1$ Higgs field, the phase diagram essentially consists of a confining and Higgs phase which are continuously connected and there exists no local gauge-invariant order parameter which distinguishes the two phases in accordance with Elitzur's theorem \cite{Elitzur:1975im}. However for $N_h\geq 2$, there may exist additional global flavor symmetries of the action which can be broken spontaneously. There may now exist a broken phase completely separated from the symmetric phase signalled by a non-vanishing gauge-invariant order parameter which breaks the corresponding global flavor symmetry. If we look at the norm of the Higgs field, which is a gauge-invariant quantity, for the case of $N_h=1$, it is always non-zero but shows a sharp rise as it crosses into the Higgs phase from the confining phase.

\begin{figure}
  \centering
   \includegraphics[width=0.33\textwidth]{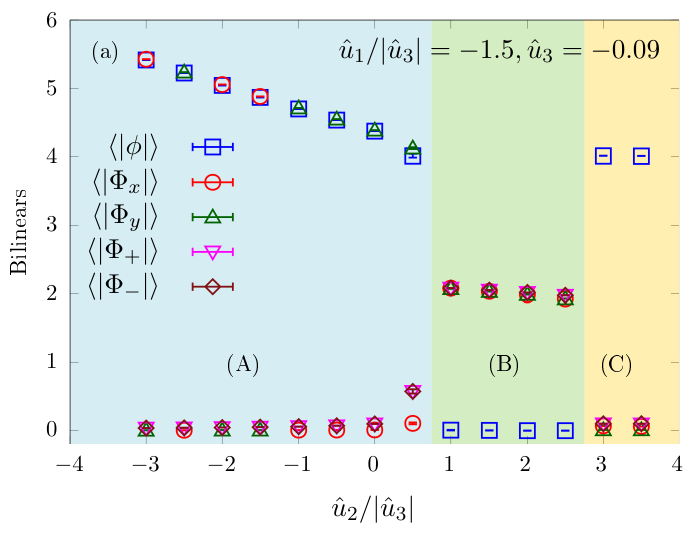}
   \includegraphics[width=0.33\textwidth]{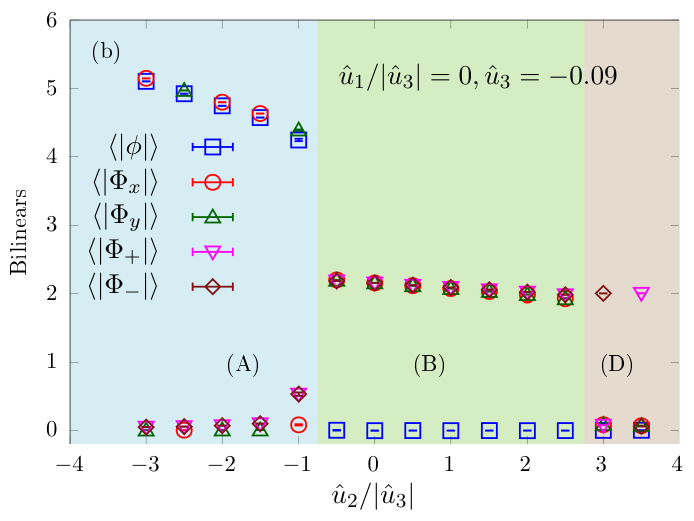}
   \includegraphics[width=0.33\textwidth]{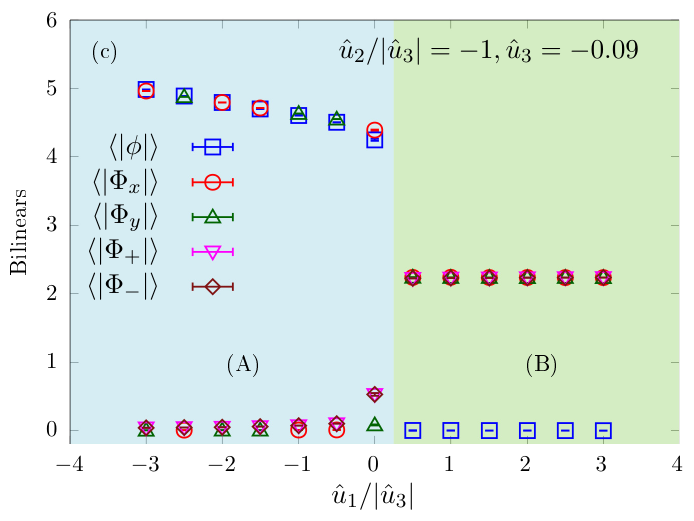}
   \includegraphics[width=0.33\textwidth]{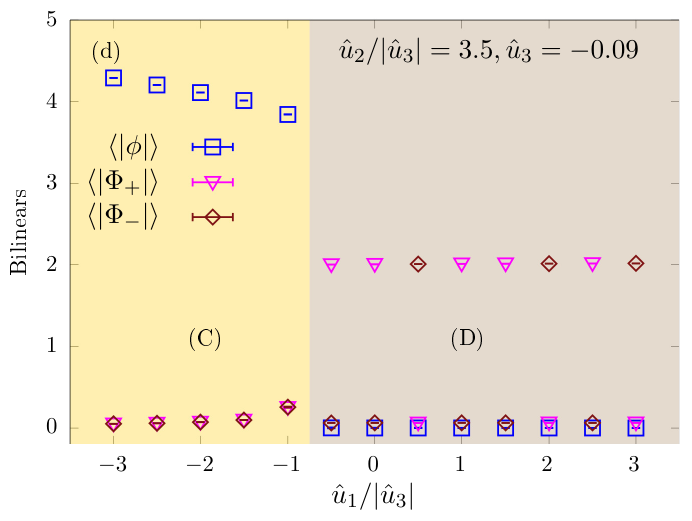}  
  \caption{Bilinears as a function of $\hat{u}_1$ or $\hat{u}_2$ along the Higgs phase diagram for a negative value of $\hat{u}_3=-0.09$. The results have been obtained on lattices of size $N_s=N_t=12$.}\label{fig:neg_u3}
 \end{figure}
For the $N_h=4$ theory, we study the bilinears which signal the breaking of the various global flavor symmetries in the Higgs phases. In Fig~\ref{fig:rhosq}, we show the variation of the sum of the norm of all Higgs fields, $B_{ii}$, as a function of the hopping parameter $\kappa$ at $\beta=8.0$, $\lambda=1.0$ and $\hat{u}_3=0.09$ for various fixed values of the quartic couplings $\hat{u}_1$ and $\hat{u}_2$. In the same plot, we also show the variation of the bilinear $\expval{\phi}$. For $\hat{u}_1=\hat{u}_2=0.18$, the system goes into the Higgs phase (G) around $\kappa\sim 1.5$ as we increase $\kappa$ while in the case of $\hat{u}_1=\hat{u}_2=-0.18$, it goes to phase (A) around the same value of $\kappa$. The bilinear $\expval{\phi}$ is non-zero in phase (A) and vanishes in phase (G) (\textit{cf}.~Fig.~\ref{fig:meanfld} (a)). As can be seen, the norm is always non-zero and shows a sharp change which coincides with the global symmetry breaking. 
 
 We performed our simulations in the broken Higgs regions of the phase diagram by setting $\beta=8.0$, $\kappa=3.0$ and $\lambda=1.0$. 
 The $\kappa$ and $\lambda$ values have been chosen appropriately such that the system remains in a stable Higgs phase, \textit{i.e.}, the potential is bounded from below for small values of the three quartic couplings $\hat{u}_1$, $\hat{u}_2$ and $\hat{u}_3$. The set of plots in Fig.~\ref{fig:pos_u3} describe the phase transitions among the various phases at different values of the quartic couplings $\hat{u}_1$ and $\hat{u}_2$ for a fixed $\hat{u}_3>0$. They correspond to horizontal (fixed $u_2$) or vertical (fixed $u_1$) scans along the phase diagram depicted {in} Fig.~\ref{fig:meanfld} (a). We confirm the existence of the various phases as conjectured in mean-field theory based on the values of the bilinear observables.

 In Fig.~\ref{fig:neg_u3}, we display the $\hat{u}_1$ and $\hat{u}_2$ scans in the broken Higgs region of the phase diagram at a negative value of $\hat{u}_3=-0.09$. Again, we find qualitatively the corresponding Higgs phases as predicted in Fig.~\ref{fig:meanfld} (b). The locations of the phase transitions show obvious deviations from the mean-field predictions. 

\begin{figure}
     \centering
     \includegraphics[width=0.43\textwidth]{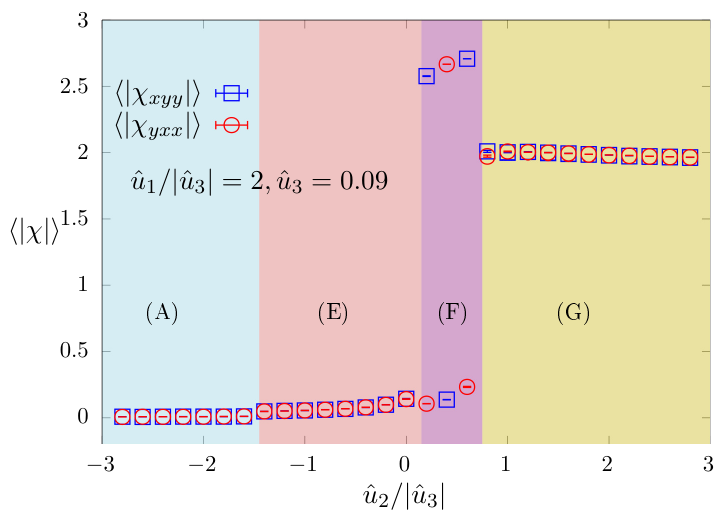}
     \caption{Variation of the time-reversal symmetry breaking order parameters, $\expval{|\chi_{xyy}|}$ and $\expval{|\chi_{yxx}|}$, as a function of $\hat{u}_2/|\hat{u}_3|$ at fixed $\hat{u}_1/|\hat{u}_3=2|$ and $\hat{u}_3=0.09$. The results have been obtained on $N_s=N_t=16$ lattices.}
     \label{fig:chi}
 \end{figure}

 Next, we looked at the expectation values of  the global $\mathbb{Z}_2$-symmetry breaking order parameters, $\chi_{xyy}$ and $\chi_{yxx}$, defined in Eqs.~{(\ref{Eq:chixyy})} and {(\ref{Eq:chiyxx})}, which signal time-reversal symmetry breaking in the {cuprate system}. In Fig.~\ref{fig:chi}, we show the variation of the absolute values, $\expval{|\chi_{xyy}|}$ and $\expval{|\chi_{yxx}|}$ as a function of the quartic coupling $\hat{u}_2$ at fixed $\hat{u}_1$ and $\hat{u}_3$. The system undergoes transitions from phases (A) to (E), (E) to (F) and (F) to (G) in the plot range. As expected, the observables are non-zero only in the phases (F) and (G), with either being non-zero in (F) and both non-zero in (G).

\begin{figure}
     \centering
     \includegraphics[width=0.42\textwidth]{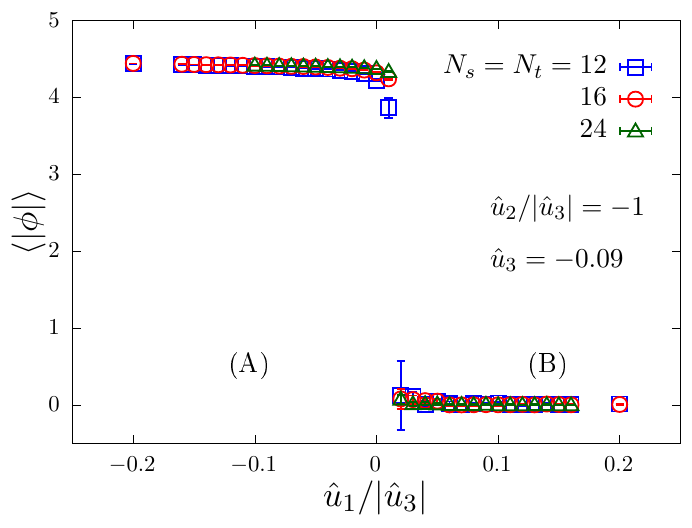}
     \includegraphics[width=0.42\textwidth]{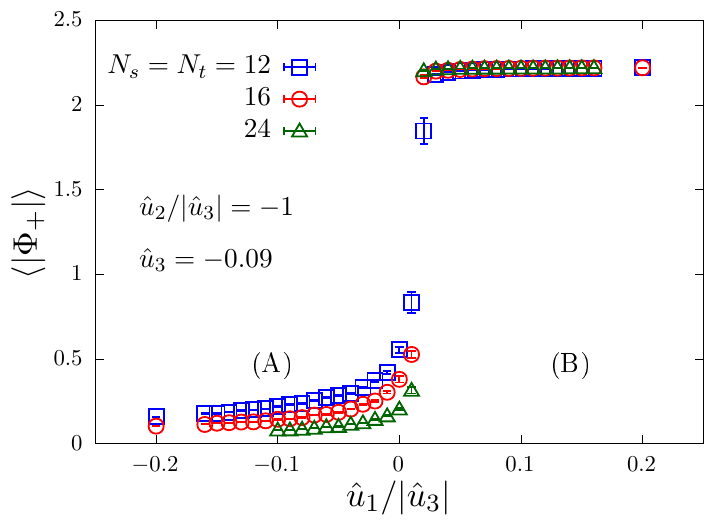}
     \caption{Volume dependence of the bilinears, $\expval{|\phi|}$ and $\expval{|\Phi_+|}$ as a function of $\hat{u}_1/|\hat{u}_3|$ across the transition between phases (A) and (B) at $\hat{u}_2/|\hat{u}_3|=-1$ and $\hat{u}_3=-0.09$.}
     \label{fig:voldep}
 \end{figure}

 In Fig.~\ref{fig:voldep}, we demonstrate the volume dependence of bilinears, $\expval{|\phi|}$ and $\expval{|\Phi_+|}$, along the transition between phases (A) and (B) as a function of $\hat{u}_1$. There appears to be an overall small dependence on the volume for different observables. Hence, the majority of our results for the phase diagram have been obtained for $N_s=N_t=12$ lattices.

\subsection{Study of confining-deconfining transitions with Polyakov loop}
\begin{figure}[t]
     \centering
     \includegraphics[width=0.43\textwidth]{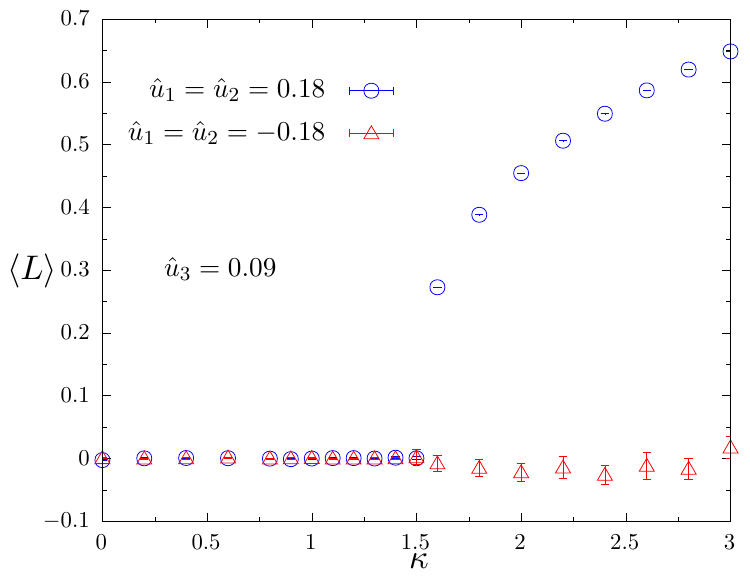}
     \caption{Polyakov loop plotted as a function of $\kappa$ for various values of $\hat{u}_1$ and $\hat{u}_2$ at fixed $\hat{u}_3=0.09$. The results have been obtained on $N_s=N_t=16$ lattices.}
     \label{fig:ploopkappa}
 \end{figure}

 Next, we study the Polyakov loop, $L$, in the different phases of the theory to understand the confining properties of particles having a fundamental gauge charge. In the symmetric phase of the $SU(2)$ gauge theory with adjoint Higgs fields, the system is expected to be in the confining phase. Unlike the case of fundamental Higgs fields, the adjoint fields do not explicitly break the global center symmetry of the gauge group, in this case, $\mathbb{Z}_2$, and therefore, the Polyakov loop should be zero in the confining phase. As an example, the variation of $\expval{L}$ as a function of the hopping parameter $\kappa$ is illustrated in Fig.~\ref{fig:ploopkappa} which clearly demonstrates that until the transition to the Higgs phase at $\kappa \lesssim 1.5$ for $\beta=8.0,\lambda=1.0$, and $\hat{u}_3=0.09$, the system is confining.

\begin{figure}[t]
     \centering
     \includegraphics[width=0.45\textwidth]{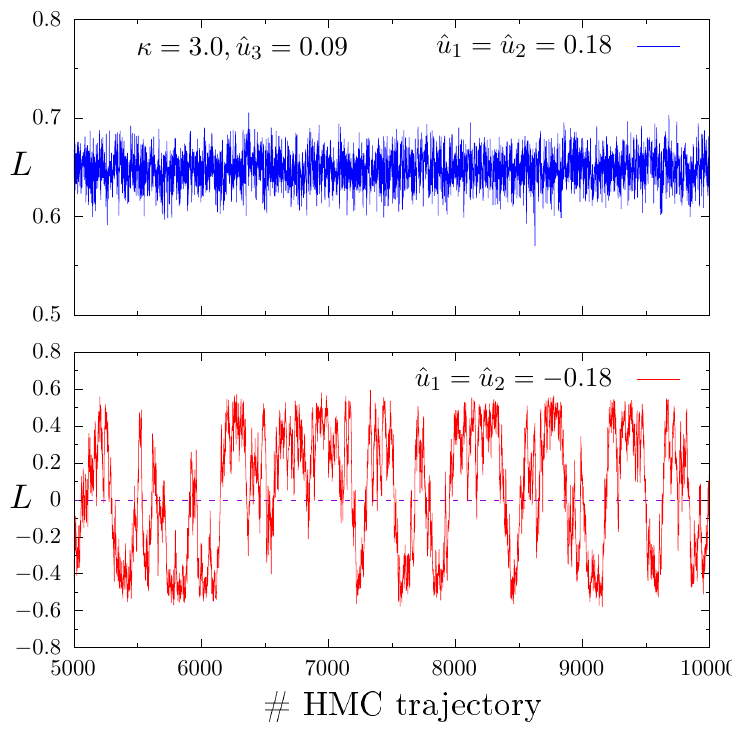}
     \includegraphics[width=0.44\textwidth]{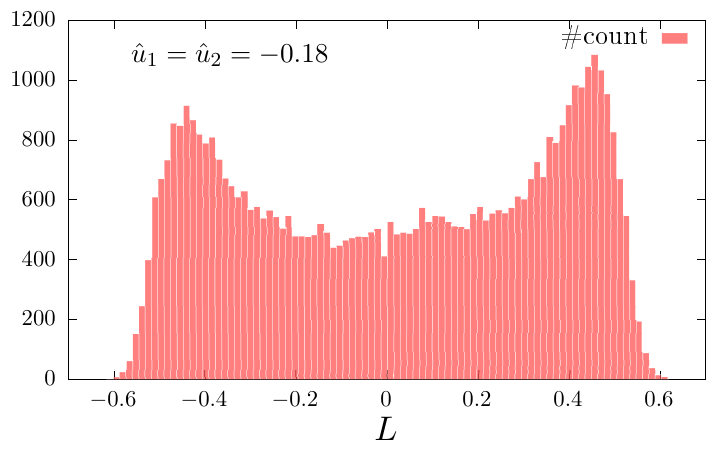}
     \caption{Polyakov loop time history obtained for - \textit{Top}: $\hat{u}_1/\hat{u}_3=\hat{u}_2/\hat{u}_3=2$ and \textit{Middle}: $\hat{u}_1/\hat{u}_3=\hat{u}_2/\hat{u}_3=-2$ at fixed $\beta=8.0$, $\kappa=3.0$, $\lambda=1.0$ and $\hat{u}_3=0.09$ on a $16^3$ lattice. \textit{Bottom}: Histogram of the $L$ values of the HMC trajectory displayed in the middle plot, \textit{i.e.}, for $\hat{u}_1/\hat{u}_3=\hat{u}_2/\hat{u}_3=-2$. The results have been obtained on $N_s=N_t=16$ lattices.}
     \label{fig:ploopTH}
 \end{figure}

 An interesting phenomenon of the Polyakov loop in the Higgs phases is visible in Fig.~\ref{fig:ploopkappa}. For quartic couplings $\hat{u}_1/\hat{u}_3=\hat{u}_2/\hat{u}_3=2$, the system is in the Higgs phase (G) for $\kappa \gtrsim 1.5$, whereas it is in the Higgs phase (A) for $\hat{u}_1/\hat{u}_3=\hat{u}_2/\hat{u}_3=-2$ (\textit{cf}.~Fig.~\ref{fig:pos_u3}). In phase (G), the Polyakov loop $L$ is non-zero and fluctuates around one of the two degenerate vacua in a given HMC run as seen in the time history plot in Fig.~\ref{fig:ploopTH} top. However, in phase (A), $L$ fluctuates between the two vacua as seen in the time history plot and the corresponding histogram in Fig.~\ref{fig:ploopTH} middle and bottom respectively. Thus, the resulting expectation value $\expval{L}\sim 0$, as seen in Fig.~\ref{fig:ploopkappa}, does not automatically mean that the system is confining. One should therefore look at the Monte Carlo time history and the absolute value $\expval{|L|}$.
 
\begin{figure}[t]
     \centering
     \includegraphics[width=0.43\textwidth]{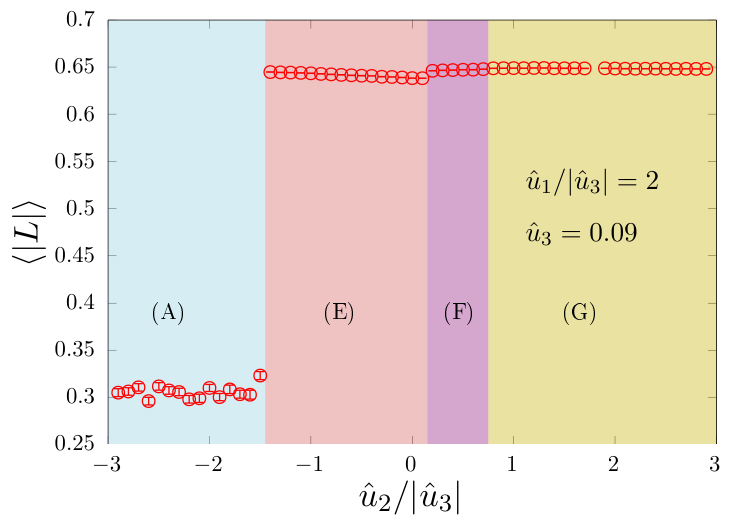}
     \caption{Absolute value of the Polyakov loop as a function of $\hat{u}_2/|\hat{u}_3|$ obtained at fixed $\hat{u}_1/|\hat{u}_3|=2$ and $\hat{u}_3=0.09$. The results have been obtained on $N_s=N_t=16$ lattices.}
     \label{fig:absploop}
\end{figure}

 In our simulations in the Higgs phases, we find that the Polyakov loop, $L$, has a finite value in the phases (C),(D),(E) and (F), all of which have the same gauge symmetry breaking pattern, $SU(2)\to \mathbb{Z}_2$. In the $SU(2)\to U(1)$ broken Higgs phases (A) and (B), we observe $\expval{L}\sim 0$  but the ensemble average of the absolute value $\expval{|L|}\neq 0$. In Fig.~\ref{fig:absploop}, we describe an instance of $\expval{|L|}$ as a function of the quartic coupling $\hat{u}_2$ keeping all other couplings fixed. The bilinear observables and $\mathbb{Z}_2$ symmetry breaking observables ($\chi_{xyy}$ and $\chi_{yxx}$) for the same values of couplings have been demonstrated in Fig.~\ref{fig:pos_u3} (d) and \ref{fig:chi} respectively, which depict the transitions from phases (A)$\rightarrow$(E)$\rightarrow$(F)$\rightarrow$(G) from left to right. The phase (A) with remnant $U(1)$ symmetry below $\hat{u}_2/|\hat{u}_3|\sim -1.5$ has $\expval{|L|} \neq 0$ which results from the oscillatory behavior of the Polyakov loop between two vacua.

\begin{figure}[t]
     \centering
     \includegraphics[width=0.43\textwidth]{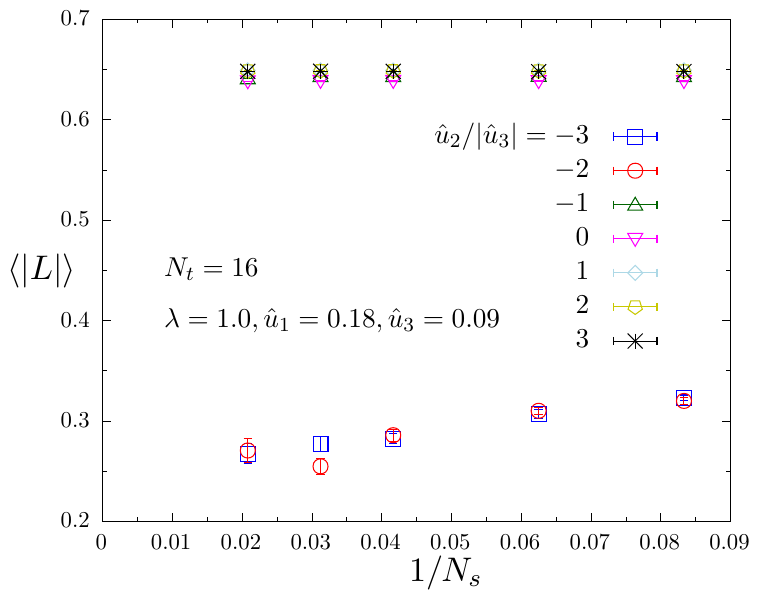}
     \includegraphics[width=0.43\textwidth]{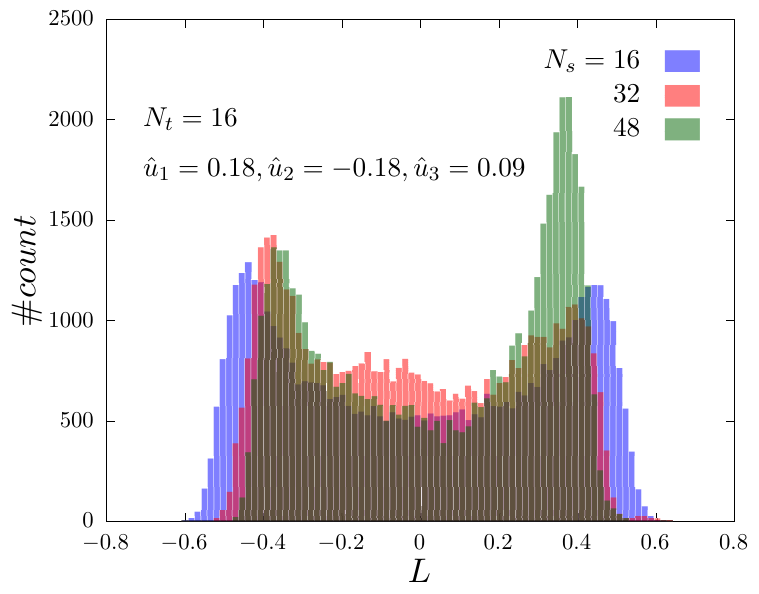}
     \caption{\textit{Top}: Absolute value of Polyakov loop as a function of the inverse spatial extent $N_s$ for fixed couplings and $N_t=16$. \textit{Bottom}: Histogram of the Polyakov loop values obtained in MC runs for three spatial lattice volumes at fixed $N_t=16$, corresponding to $\hat{u}_2/|\hat{u}_3|=-2$ shown above. All of the above results have been obtained for fixed couplings, $\beta=8.0, \kappa=3.0$ and $\lambda=1.0$.}
     \label{fig:absploopNs}
\end{figure}

The deconfinement phase coinciding with the $SU(2)\to U(1)$ Higgs phases appears to be distinctly different from the Higgs phases with remnant $\mathbb{Z}_2$ symmetry. In the former case, the Polyakov loop potential seems to have a shallower double well structure barrier leading to frequent tunneling between the two vacua. To investigate this further, we have studied the spatial volume and temporal width dependence of the Polyakov loop.
In Fig.~\ref{fig:absploopNs}, we show the variation of the absolute value of the Polyakov loop as the spatial system size $N_s$ is taken to infinity for a fixed temporal extent $N_t=16$. The data points for $\hat{u}_2/|\hat{u}_3|>-2$ correspond to the $SU(2)\to \mathbb{Z}_2$ phases (E, F, G) and $\hat{u}_2/|\hat{u}_3|=-2,-3$ to the $SU(2)\to U(1)$ phase (A). However, the $L$ itself has the same oscillatory behavior between the two degenerate vacua as a function of the Monte Carlo time in the $U(1)$-symmetric phases as demonstrated in the histogram in Fig.~\ref{fig:absploopNs} bottom. The $\mathbb{Z}_2$ symmetric phase, on the other hand, is a stable deconfined phase as already evidenced by the top plot in Fig.~\ref{fig:ploopTH}.

\begin{figure}
     \centering
     \includegraphics[width=0.43\textwidth]{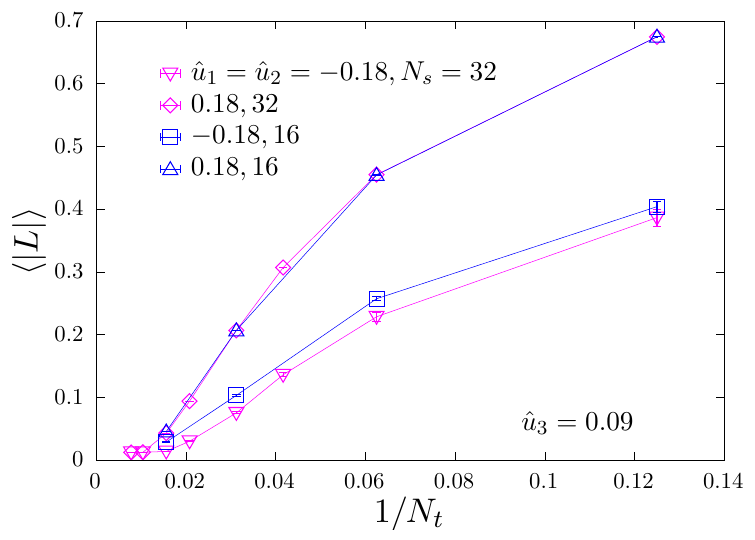}
     \caption{Absolute value of Polyakov loop as a function of the inverse temporal extent $N_t$ for fixed couplings and $N_s=16$ and $32$. All of the above results have been obtained for fixed couplings, $\beta=8.0, \kappa=3.0$ and $\lambda=1.0$.}
     \label{fig:absploopNt}
 \end{figure}

 In Fig.~\ref{fig:absploopNt}, we display the dependence of $\expval{|L|}$ on the temporal extent $N_t$ for fixed spatial dimension $N_s=16$ and $32$. As $N_t$ increases, the circumference of the Polyakov loop becomes larger in units of the lattice spacing, and its value goes to zero at finite $N_t$, \textit{i.e.}, at finite temperature for both $U(1)$ and $\mathbb{Z}_2$ symmetric Higgs phases.

 In \cite{Sachdev:2018nbk}, it has been argued that the broken phase with remnant $U(1)$ symmetry would ultimately be confining at large length scales due to the proliferation of monopoles. In our studies, we find this phase to be consistent with a deconfining phase in the finite volumes that we accessed. However, the Polyakov loop potential appears to be quite flat, leading to frequent tunneling between the two vacua. With the increase of spatial volume, we do not find any indication of departure from this behavior. It might be possible that our volumes have to be larger than some critical value beyond which confinement sets in, as was found in \cite{Davis:2001mg} for the $d=3$ $SU(2)$ theory with a single adjoint Higgs. On the other hand, we find the $\mathbb{Z}_{2}$ phases to be deconfining with a Polyakov loop that fluctuates only in one of the vacua. With increasing temporal extent or decreasing temperature in units of the lattice spacing, the Polyakov loop vanishes as expected for both the phases.
\section{Conclusion}
\label{sec:conclusion}
We have studied the $3-$dimensional $SU(2)$ gauge theory coupled to $N_h=4$ adjoint Higgs fields on the lattice. The theory is phenomenologically important as an effective theory for the pseudogap phase in cuprate superconductors. We find that the phase diagram of the theory shows a rich variety of Higgs phases, where, in the perturbative description, the $SU(2)$ symmetry is broken down either to $U(1)$ or $\mathbb{Z}_2$. The Higgs phases can be characterized by a set of gauge-invariant bilinears constructed out of the $N_h=4$ adjoint Higgs fields. At a single value of the gauge coupling, corresponding to $\beta=8.0$, we find the phase structure to be qualitatively consistent with mean-field predictions. In a future study, we aim to investigate the $\beta$ dependence of the Higgs phase diagram and the nature of the phase transitions.

We have further studied the Polyakov loop to investigate the confinement properties of the various Higgs phases. We find that the broken phases with remnant $U(1)$ or $\mathbb{Z}_2$ symmetry are deconfining with non-zero values of the Polyakov loop. However, the $U(1)$ phase displays a Polyakov loop which rapidly tunnels between the two minima of the Polyakov loop potential during a Monte-Carlo run. The behavior appears to be unchanged with increasing spatial volumes. It would be interesting to study whether this behavior is correlated with the presence of monopoles in the Higgs phase with remnant $U(1)$ symmetry. The remnant $U(1)$-symmetric phase may turn out be confining, as expected, if we are able to access yet larger volumes, which we plan to do in the future.

\acknowledgments
The authors thank Subir Sachdev and Mathias Scheurer for very useful discussions.
This work is supported by NSTC grants 110-2112-M-002-034-MY3, 112-2112-M-A49-021-MY3 and 113-2639-M-002-006-ASP.  Numerical computations for this work were carried out on the HPC facilities at National Taiwan University and National Yang Ming Chiao Tung University, as well as on the ASGC resources at Academia Sinica.  
A.R and G.C. acknowledge financial support from CSIC, Acciones bilaterales CSIC-NSTC (BILTW22007).
This work is also funded by the European Union's Horizon Europe Framework Programme (HORIZON) under the ERA Chair scheme with grant agreement no.\ 101087126 and the Ministry of Science, Research and Culture of the State of Brandenburg within the Centre for Quantum Technologies and Applications (CQTA).

\begin{center}
    \includegraphics[width = 0.05\textwidth]{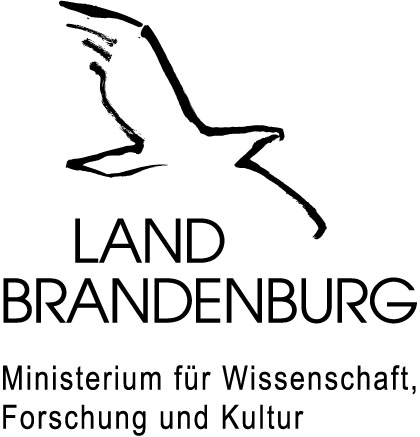}
\end{center}

\appendix
\section{Flavor symmetries of the Higgs potential} \label{app:symm}
The Higgs potential is symmetric under various discrete and continuous flavor symmetries, which correspond to the following symmetries of the physical system. We use the notation of Ref.~\cite{Sachdev:2018nbk}.

\begin{enumerate}
    \item \textit{Time-reversal symmetry}: This is a global $\mathbb{Z}_2$ flavor symmetry which transforms the Higgs fields $\Phi_l \to -\Phi_l$ for $l=1$--4. All the bilinear observables in Eq.~{(\ref{Eq:orderparams})} are invariant under this transformation but the observables $\chi_{xyy}$ and $\chi_{yxx}$ break this symmetry.
    \item \textit{Translations on the square lattice}: The translations by a square lattice vector $\bvec{\eta}=(\eta_x,\eta_y)$ correspond to the following phase factor being multiplied to the Higgs fields,
    \begin{equation}
        \begin{pmatrix}
        \Phi_1\\
        \Phi_2
        \end{pmatrix} \to \exp^{i \bvec{K}_x \cdot \bvec{\eta}} \begin{pmatrix}
        \Phi_1\\
        \Phi_2
        \end{pmatrix}\,,
    \end{equation}
    and similarly for $(\Phi_3, \Phi_4)$ with $\bvec{K}_y$, where $\bvec{K}_{x/y}$ are the incommensurate wave vectors defined {below Eq.~(\ref{Eq:orderparams})}. Since the wave vectors are incommensurate, the phases densely cover $U(1)$ resulting in a $U(1)\times U(1)$ flavor symmetry.
    \item \textit{$C_4$ fourfold rotation along $z$ axis}: Under this transformation, the Higgs fields are rotated in the flavor space in the following manner, $(\Phi_1, \Phi_2, \Phi_3, \Phi_4)$ $\to$ $(\Phi_3, \Phi_4, \Phi_1, -\Phi_2)$. 
    \item \textit{$\sigma_{x}$ ($\sigma_{y}$) reflection at the $xz$ ($yz$) plane}: Under these discrete transformations, the Higgs fields transform as $(\Phi_1, \Phi_2, \Phi_3, \Phi_4)$ $\to$ $(\Phi_1, \Phi_2, \Phi_3, -\Phi_4)$ or $(\Phi_1, -\Phi_2, \Phi_3, \Phi_4)$.
\end{enumerate}
\onecolumngrid
\section{Details of HMC}\label{app:hmc}
For the HMC update of the $N_h=4$ adjoint Higgs fields 
 $\Phi_n(x) = \Phi_n^{\alpha}(x)\tau_\alpha$ 
 ($\tau_\alpha = \sigma_\alpha/2$, $n=1,\ldots,N_h$), 
 we introduce the 
 conjugate momenta $\pi_n(x) = \pi_n^{\alpha}(x)\tau_\alpha$ in 
 the Molecular Dynamics (MD) Hamiltonian as
 \begin{equation}
 H = S + \sum_{x}\sum_{n=1}^{N_h} \Tr \pi_n({x})\pi_n({x}) = S + \frac{1}{2}\sum_{x}\sum_{n=1}^{N_h}\sum_{\alpha=1}^3 \pi_n^{\alpha}(x)\pi_n^{\alpha}(x) \, ,
 \end{equation}
 where $S$ is the lattice action given in Eq.~(\ref{Eq:lattaction}).
 The MD evolution equations are
 \begin{align}
 \dot\Phi_n^{\alpha}(x) = \pi_n^{\alpha}(x)\;, \quad
 \dot{\pi}_n^{\alpha}(x) = -\frac{\partial S}{\partial \Phi_n^{\alpha}(x)} \,.
 \end{align} 
 
 The force terms corresponding to various terms of the Higgs potential are summarized below. The shorthand notation $B_{mn}(x) = \Phi^\alpha_m(x) \Phi^\alpha_n(x)$ has been used.

 \underline{$\bm{\lambda}$ terms}
 
 \begin{align}
  S_\lambda &= \lambda \sum_x \Big( \sum_{n=1}^{N_h} \left( B_{nn} - 1\right)^2 + \sum_{n \neq m}^{N_h} B_{nn}B_{mm} \Big) \\
  \frac{\partial S_\lambda}{\partial \Phi_n^{\alpha}(x)} &= 4\lambda\sum_{n=1}^{N_h}\left( B_{nn}(x)-1\right)\Phi_n^{\alpha}(x) + 4\lambda \sum_{m\neq n}^{N_h}B_{mm}(x) \Phi_n^{\alpha}(x)
 \end{align}
 
 \underline{$\bm{u_1}$ terms}
 \begin{equation}
 S_{u_1} =\frac{\hat{u_1}}{2} \sum_x \Bigg[ \frac{1}{2}\sum_{n=1}^{N_h} (B_{nn})^2 
  + B_{11}B_{22} + B_{33}B_{44} - B_{11}B_{33} - B_{11}B_{44} - B_{22}B_{33} - B_{22}B_{44} \Bigg]
  \end{equation}
  \begin{align}
  \frac{\partial S_{u_1}}{\partial \Phi_n^{\alpha}(x)} =\ &\hat{u}_1 \Bigg[ B_{nn}(x)\Phi_n^\alpha(x) + \left\lbrace B_{22}(x) - B_{33}(x) - B_{44}(x)\right\rbrace\Phi_1^\alpha(x) \delta_{1n} \nonumber \\
  &+ \left\lbrace B_{11}(x) - B_{33}(x) - B_{44}(x)\right\rbrace\Phi_2^\alpha(x) \delta_{2n} \nonumber \\
  &+ \left\lbrace B_{44}(x) - B_{11}(x) - B_{22}(x)\right\rbrace\Phi_3^\alpha(x) \delta_{3n} 
  + \left\lbrace B_{33}(x) - B_{11}(x) - B_{22}(x)\right\rbrace\Phi_4^\alpha(x) \delta_{4n}
  \Bigg]
 \end{align}
 
 \underline{$\bm{u_2}$ terms}
 \begin{equation}
 S_{u_2} = \hat{u_2} \sum_x \Bigg[\frac{1}{2}\sum_{n=1}^{N_h} (B_{nn})^2 
  + 2(B_{12})^2 + 2(B_{34})^2 - B_{11}B_{22} - B_{33}B_{44} \Bigg]
 \end{equation}
 \begin{align}
 \frac{\partial S_{u_2}}{\partial \Phi_n^{\alpha}(x)} = 2\hat{u_2}&  \Bigg[ B_{nn}(x)\Phi_n^\alpha(x) + 2 B_{12}(x)\left\lbrace \Phi_1^\alpha(x) \delta_{2n} + \Phi_2^\alpha(x) \delta_{1n}\right\rbrace \quad+ 2 B_{34}(x)\left\lbrace \Phi_3^\alpha(x) \delta_{4n} + \Phi_4^\alpha(x) \delta_{3n}\right\rbrace \nonumber \\
 &\quad- B_{11}(x)\Phi_2^\alpha(x) \delta_{2n} - B_{22}(x)\Phi_1^\alpha(x) \delta_{1n} \quad- B_{33}(x)\Phi_4^\alpha(x) \delta_{4n} - B_{44}(x)\Phi_3^\alpha(x) \delta_{3n}
 \Bigg]
 \end{align}
 
 \underline{$\bm{u_3}$ terms}
 \begin{align}
 S_{u_3} &= \hat{u_3} \sum_x \sum_{n \neq m}^{N_h} B_{mn}B_{mn} \\
 \frac{\partial S_{u_2}}{\partial \Phi_n^{\alpha}(x)} &= 4\hat{u_3}\sum_{m\neq n}^{N_h} B_{mn}(x) \Phi_n^\alpha (x)
 \end{align}
\twocolumngrid

\bibliography{main} 

\begin{thebibliography}{33}%
\makeatletter
\providecommand \@ifxundefined [1]{%
 \@ifx{#1\undefined}
}%
\providecommand \@ifnum [1]{%
 \ifnum #1\expandafter \@firstoftwo
 \else \expandafter \@secondoftwo
 \fi
}%
\providecommand \@ifx [1]{%
 \ifx #1\expandafter \@firstoftwo
 \else \expandafter \@secondoftwo
 \fi
}%
\providecommand \natexlab [1]{#1}%
\providecommand \enquote  [1]{``#1''}%
\providecommand \bibnamefont  [1]{#1}%
\providecommand \bibfnamefont [1]{#1}%
\providecommand \citenamefont [1]{#1}%
\providecommand \href@noop [0]{\@secondoftwo}%
\providecommand \href [0]{\begingroup \@sanitize@url \@href}%
\providecommand \@href[1]{\@@startlink{#1}\@@href}%
\providecommand \@@href[1]{\endgroup#1\@@endlink}%
\providecommand \@sanitize@url [0]{\catcode `\\12\catcode `\$12\catcode
  `\&12\catcode `\#12\catcode `\^12\catcode `\_12\catcode `\%12\relax}%
\providecommand \@@startlink[1]{}%
\providecommand \@@endlink[0]{}%
\providecommand \url  [0]{\begingroup\@sanitize@url \@url }%
\providecommand \@url [1]{\endgroup\@href {#1}{\urlprefix }}%
\providecommand \urlprefix  [0]{URL }%
\providecommand \Eprint [0]{\href }%
\providecommand \doibase [0]{https://doi.org/}%
\providecommand \selectlanguage [0]{\@gobble}%
\providecommand \bibinfo  [0]{\@secondoftwo}%
\providecommand \bibfield  [0]{\@secondoftwo}%
\providecommand \translation [1]{[#1]}%
\providecommand \BibitemOpen [0]{}%
\providecommand \bibitemStop [0]{}%
\providecommand \bibitemNoStop [0]{.\EOS\space}%
\providecommand \EOS [0]{\spacefactor3000\relax}%
\providecommand \BibitemShut  [1]{\csname bibitem#1\endcsname}%
\let\auto@bib@innerbib\@empty
\bibitem [{\citenamefont {Salam}\ and\ \citenamefont
  {Ward}(1959)}]{Salam:1959zz}%
  \BibitemOpen
  \bibfield  {author} {\bibinfo {author} {\bibfnamefont {A.}~\bibnamefont
  {Salam}}\ and\ \bibinfo {author} {\bibfnamefont {J.~C.}\ \bibnamefont
  {Ward}},\ }\bibfield  {title} {\bibinfo {title} {{Weak and electromagnetic
  interactions}},\ }\href {https://doi.org/10.1007/BF02726525} {\bibfield
  {journal} {\bibinfo  {journal} {Nuovo Cim.}\ }\textbf {\bibinfo {volume}
  {11}},\ \bibinfo {pages} {568} (\bibinfo {year} {1959})}\BibitemShut
  {NoStop}%
\bibitem [{\citenamefont {Glashow}(1961)}]{Glashow:1961tr}%
  \BibitemOpen
  \bibfield  {author} {\bibinfo {author} {\bibfnamefont {S.~L.}\ \bibnamefont
  {Glashow}},\ }\bibfield  {title} {\bibinfo {title} {{Partial Symmetries of
  Weak Interactions}},\ }\href {https://doi.org/10.1016/0029-5582(61)90469-2}
  {\bibfield  {journal} {\bibinfo  {journal} {Nucl. Phys.}\ }\textbf {\bibinfo
  {volume} {22}},\ \bibinfo {pages} {579} (\bibinfo {year} {1961})}\BibitemShut
  {NoStop}%
\bibitem [{\citenamefont {Weinberg}(1967)}]{Weinberg:1967tq}%
  \BibitemOpen
  \bibfield  {author} {\bibinfo {author} {\bibfnamefont {S.}~\bibnamefont
  {Weinberg}},\ }\bibfield  {title} {\bibinfo {title} {{A Model of Leptons}},\
  }\href {https://doi.org/10.1103/PhysRevLett.19.1264} {\bibfield  {journal}
  {\bibinfo  {journal} {Phys. Rev. Lett.}\ }\textbf {\bibinfo {volume} {19}},\
  \bibinfo {pages} {1264} (\bibinfo {year} {1967})}\BibitemShut {NoStop}%
\bibitem [{\citenamefont {Bardeen}\ \emph
  {et~al.}(1957{\natexlab{a}})\citenamefont {Bardeen}, \citenamefont {Cooper},\
  and\ \citenamefont {Schrieffer}}]{Bardeen:1957kj}%
  \BibitemOpen
  \bibfield  {author} {\bibinfo {author} {\bibfnamefont {J.}~\bibnamefont
  {Bardeen}}, \bibinfo {author} {\bibfnamefont {L.~N.}\ \bibnamefont
  {Cooper}},\ and\ \bibinfo {author} {\bibfnamefont {J.~R.}\ \bibnamefont
  {Schrieffer}},\ }\bibfield  {title} {\bibinfo {title} {{Microscopic theory of
  superconductivity}},\ }\href {https://doi.org/10.1103/PhysRev.106.162}
  {\bibfield  {journal} {\bibinfo  {journal} {Phys. Rev.}\ }\textbf {\bibinfo
  {volume} {106}},\ \bibinfo {pages} {162} (\bibinfo {year}
  {1957}{\natexlab{a}})}\BibitemShut {NoStop}%
\bibitem [{\citenamefont {Bardeen}\ \emph
  {et~al.}(1957{\natexlab{b}})\citenamefont {Bardeen}, \citenamefont {Cooper},\
  and\ \citenamefont {Schrieffer}}]{Bardeen:1957mv}%
  \BibitemOpen
  \bibfield  {author} {\bibinfo {author} {\bibfnamefont {J.}~\bibnamefont
  {Bardeen}}, \bibinfo {author} {\bibfnamefont {L.~N.}\ \bibnamefont
  {Cooper}},\ and\ \bibinfo {author} {\bibfnamefont {J.~R.}\ \bibnamefont
  {Schrieffer}},\ }\bibfield  {title} {\bibinfo {title} {{Theory of
  superconductivity}},\ }\href {https://doi.org/10.1103/PhysRev.108.1175}
  {\bibfield  {journal} {\bibinfo  {journal} {Phys. Rev.}\ }\textbf {\bibinfo
  {volume} {108}},\ \bibinfo {pages} {1175} (\bibinfo {year}
  {1957}{\natexlab{b}})}\BibitemShut {NoStop}%
\bibitem [{\citenamefont {Wilson}(1974)}]{Wilson:1974sk}%
  \BibitemOpen
  \bibfield  {author} {\bibinfo {author} {\bibfnamefont {K.~G.}\ \bibnamefont
  {Wilson}},\ }\bibfield  {title} {\bibinfo {title} {{Confinement of Quarks}},\
  }\href {https://doi.org/10.1103/PhysRevD.10.2445} {\bibfield  {journal}
  {\bibinfo  {journal} {Phys. Rev. D}\ }\textbf {\bibinfo {volume} {10}},\
  \bibinfo {pages} {2445} (\bibinfo {year} {1974})}\BibitemShut {NoStop}%
\bibitem [{\citenamefont {Appelquist}\ and\ \citenamefont
  {Pisarski}(1981)}]{Appelquist:1981vg}%
  \BibitemOpen
  \bibfield  {author} {\bibinfo {author} {\bibfnamefont {T.}~\bibnamefont
  {Appelquist}}\ and\ \bibinfo {author} {\bibfnamefont {R.~D.}\ \bibnamefont
  {Pisarski}},\ }\bibfield  {title} {\bibinfo {title} {{High-Temperature
  Yang-Mills Theories and Three-Dimensional Quantum Chromodynamics}},\ }\href
  {https://doi.org/10.1103/PhysRevD.23.2305} {\bibfield  {journal} {\bibinfo
  {journal} {Phys. Rev. D}\ }\textbf {\bibinfo {volume} {23}},\ \bibinfo
  {pages} {2305} (\bibinfo {year} {1981})}\BibitemShut {NoStop}%
\bibitem [{\citenamefont {Elitzur}(1975)}]{Elitzur:1975im}%
  \BibitemOpen
  \bibfield  {author} {\bibinfo {author} {\bibfnamefont {S.}~\bibnamefont
  {Elitzur}},\ }\bibfield  {title} {\bibinfo {title} {{Impossibility of
  Spontaneously Breaking Local Symmetries}},\ }\href
  {https://doi.org/10.1103/PhysRevD.12.3978} {\bibfield  {journal} {\bibinfo
  {journal} {Phys. Rev. D}\ }\textbf {\bibinfo {volume} {12}},\ \bibinfo
  {pages} {3978} (\bibinfo {year} {1975})}\BibitemShut {NoStop}%
\bibitem [{\citenamefont {Fradkin}\ and\ \citenamefont
  {Shenker}(1979)}]{Fradkin:1978dv}%
  \BibitemOpen
  \bibfield  {author} {\bibinfo {author} {\bibfnamefont {E.~H.}\ \bibnamefont
  {Fradkin}}\ and\ \bibinfo {author} {\bibfnamefont {S.~H.}\ \bibnamefont
  {Shenker}},\ }\bibfield  {title} {\bibinfo {title} {{Phase Diagrams of
  Lattice Gauge Theories with Higgs Fields}},\ }\href
  {https://doi.org/10.1103/PhysRevD.19.3682} {\bibfield  {journal} {\bibinfo
  {journal} {Phys. Rev. D}\ }\textbf {\bibinfo {volume} {19}},\ \bibinfo
  {pages} {3682} (\bibinfo {year} {1979})}\BibitemShut {NoStop}%
\bibitem [{\citenamefont {Nadkarni}(1990)}]{Nadkarni:1989na}%
  \BibitemOpen
  \bibfield  {author} {\bibinfo {author} {\bibfnamefont {S.}~\bibnamefont
  {Nadkarni}},\ }\bibfield  {title} {\bibinfo {title} {{The SU(2) Adjoint Higgs
  Model in Three-dimensions}},\ }\href
  {https://doi.org/10.1016/0550-3213(90)90491-U} {\bibfield  {journal}
  {\bibinfo  {journal} {Nucl. Phys. B}\ }\textbf {\bibinfo {volume} {334}},\
  \bibinfo {pages} {559} (\bibinfo {year} {1990})}\BibitemShut {NoStop}%
\bibitem [{\citenamefont {Hart}\ \emph {et~al.}(1997)\citenamefont {Hart},
  \citenamefont {Philipsen}, \citenamefont {Stack},\ and\ \citenamefont
  {Teper}}]{Hart:1996ac}%
  \BibitemOpen
  \bibfield  {author} {\bibinfo {author} {\bibfnamefont {A.}~\bibnamefont
  {Hart}}, \bibinfo {author} {\bibfnamefont {O.}~\bibnamefont {Philipsen}},
  \bibinfo {author} {\bibfnamefont {J.~D.}\ \bibnamefont {Stack}},\ and\
  \bibinfo {author} {\bibfnamefont {M.}~\bibnamefont {Teper}},\ }\bibfield
  {title} {\bibinfo {title} {{On the phase diagram of the SU(2) adjoint Higgs
  model in (2+1)-dimensions}},\ }\href
  {https://doi.org/10.1016/S0370-2693(97)00104-4} {\bibfield  {journal}
  {\bibinfo  {journal} {Phys. Lett. B}\ }\textbf {\bibinfo {volume} {396}},\
  \bibinfo {pages} {217} (\bibinfo {year} {1997})},\ \Eprint
  {https://arxiv.org/abs/hep-lat/9612021} {arXiv:hep-lat/9612021} \BibitemShut
  {NoStop}%
\bibitem [{\citenamefont {Kajantie}\ \emph {et~al.}(1997)\citenamefont
  {Kajantie}, \citenamefont {Laine}, \citenamefont {Rummukainen},\ and\
  \citenamefont {Shaposhnikov}}]{Kajantie:1997tt}%
  \BibitemOpen
  \bibfield  {author} {\bibinfo {author} {\bibfnamefont {K.}~\bibnamefont
  {Kajantie}}, \bibinfo {author} {\bibfnamefont {M.}~\bibnamefont {Laine}},
  \bibinfo {author} {\bibfnamefont {K.}~\bibnamefont {Rummukainen}},\ and\
  \bibinfo {author} {\bibfnamefont {M.~E.}\ \bibnamefont {Shaposhnikov}},\
  }\bibfield  {title} {\bibinfo {title} {{3-D SU(N) + adjoint Higgs theory and
  finite temperature QCD}},\ }\href
  {https://doi.org/10.1016/S0550-3213(97)00425-2} {\bibfield  {journal}
  {\bibinfo  {journal} {Nucl. Phys. B}\ }\textbf {\bibinfo {volume} {503}},\
  \bibinfo {pages} {357} (\bibinfo {year} {1997})},\ \Eprint
  {https://arxiv.org/abs/hep-ph/9704416} {arXiv:hep-ph/9704416} \BibitemShut
  {NoStop}%
\bibitem [{\citenamefont {Hart}\ and\ \citenamefont
  {Philipsen}(2000)}]{Hart:1999dj}%
  \BibitemOpen
  \bibfield  {author} {\bibinfo {author} {\bibfnamefont {A.}~\bibnamefont
  {Hart}}\ and\ \bibinfo {author} {\bibfnamefont {O.}~\bibnamefont
  {Philipsen}},\ }\bibfield  {title} {\bibinfo {title} {{The Spectrum of the
  three-dimensional adjoint Higgs model and hot SU(2) gauge theory}},\ }\href
  {https://doi.org/10.1016/S0550-3213(99)00742-7} {\bibfield  {journal}
  {\bibinfo  {journal} {Nucl. Phys. B}\ }\textbf {\bibinfo {volume} {572}},\
  \bibinfo {pages} {243} (\bibinfo {year} {2000})},\ \Eprint
  {https://arxiv.org/abs/hep-lat/9908041} {arXiv:hep-lat/9908041} \BibitemShut
  {NoStop}%
\bibitem [{\citenamefont {'t~Hooft}(1974)}]{tHooft:1974kcl}%
  \BibitemOpen
  \bibfield  {author} {\bibinfo {author} {\bibfnamefont {G.}~\bibnamefont
  {'t~Hooft}},\ }\bibfield  {title} {\bibinfo {title} {{Magnetic Monopoles in
  Unified Gauge Theories}},\ }\href
  {https://doi.org/10.1016/0550-3213(74)90486-6} {\bibfield  {journal}
  {\bibinfo  {journal} {Nucl. Phys. B}\ }\textbf {\bibinfo {volume} {79}},\
  \bibinfo {pages} {276} (\bibinfo {year} {1974})}\BibitemShut {NoStop}%
\bibitem [{\citenamefont {Polyakov}(1974)}]{Polyakov:1974ek}%
  \BibitemOpen
  \bibfield  {author} {\bibinfo {author} {\bibfnamefont {A.~M.}\ \bibnamefont
  {Polyakov}},\ }\bibfield  {title} {\bibinfo {title} {{Particle Spectrum in
  Quantum Field Theory}},\ }\href@noop {} {\bibfield  {journal} {\bibinfo
  {journal} {JETP Lett.}\ }\textbf {\bibinfo {volume} {20}},\ \bibinfo {pages}
  {194} (\bibinfo {year} {1974})}\BibitemShut {NoStop}%
\bibitem [{\citenamefont {Polyakov}(1977)}]{Polyakov:1976fu}%
  \BibitemOpen
  \bibfield  {author} {\bibinfo {author} {\bibfnamefont {A.~M.}\ \bibnamefont
  {Polyakov}},\ }\bibfield  {title} {\bibinfo {title} {{Quark Confinement and
  Topology of Gauge Groups}},\ }\href
  {https://doi.org/10.1016/0550-3213(77)90086-4} {\bibfield  {journal}
  {\bibinfo  {journal} {Nucl. Phys. B}\ }\textbf {\bibinfo {volume} {120}},\
  \bibinfo {pages} {429} (\bibinfo {year} {1977})}\BibitemShut {NoStop}%
\bibitem [{\citenamefont {Mandelstam}(1976)}]{Mandelstam:1974pi}%
  \BibitemOpen
  \bibfield  {author} {\bibinfo {author} {\bibfnamefont {S.}~\bibnamefont
  {Mandelstam}},\ }\bibfield  {title} {\bibinfo {title} {{Vortices and Quark
  Confinement in Nonabelian Gauge Theories}},\ }\href
  {https://doi.org/10.1016/0370-1573(76)90043-0} {\bibfield  {journal}
  {\bibinfo  {journal} {Phys. Rept.}\ }\textbf {\bibinfo {volume} {23}},\
  \bibinfo {pages} {245} (\bibinfo {year} {1976})}\BibitemShut {NoStop}%
\bibitem [{\citenamefont {Davis}\ \emph {et~al.}(2002)\citenamefont {Davis},
  \citenamefont {Hart}, \citenamefont {Kibble},\ and\ \citenamefont
  {Rajantie}}]{Davis:2001mg}%
  \BibitemOpen
  \bibfield  {author} {\bibinfo {author} {\bibfnamefont {A.~C.}\ \bibnamefont
  {Davis}}, \bibinfo {author} {\bibfnamefont {A.}~\bibnamefont {Hart}},
  \bibinfo {author} {\bibfnamefont {T.~W.~B.}\ \bibnamefont {Kibble}},\ and\
  \bibinfo {author} {\bibfnamefont {A.}~\bibnamefont {Rajantie}},\ }\bibfield
  {title} {\bibinfo {title} {{The Monopole mass in the three-dimensional
  Georgi-Glashow model}},\ }\href {https://doi.org/10.1103/PhysRevD.65.125008}
  {\bibfield  {journal} {\bibinfo  {journal} {Phys. Rev. D}\ }\textbf {\bibinfo
  {volume} {65}},\ \bibinfo {pages} {125008} (\bibinfo {year} {2002})},\
  \Eprint {https://arxiv.org/abs/hep-lat/0110154} {arXiv:hep-lat/0110154}
  \BibitemShut {NoStop}%
\bibitem [{\citenamefont {Niemi}\ \emph {et~al.}(2023)\citenamefont {Niemi},
  \citenamefont {Rummukainen}, \citenamefont {Sepp\"a},\ and\ \citenamefont
  {Weir}}]{Niemi:2022bjg}%
  \BibitemOpen
  \bibfield  {author} {\bibinfo {author} {\bibfnamefont {L.}~\bibnamefont
  {Niemi}}, \bibinfo {author} {\bibfnamefont {K.}~\bibnamefont {Rummukainen}},
  \bibinfo {author} {\bibfnamefont {R.}~\bibnamefont {Sepp\"a}},\ and\ \bibinfo
  {author} {\bibfnamefont {D.~J.}\ \bibnamefont {Weir}},\ }\bibfield  {title}
  {\bibinfo {title} {{Infrared physics of the 3D SU(2) adjoint Higgs model at
  the crossover transition}},\ }\href {https://doi.org/10.1007/JHEP02(2023)212}
  {\bibfield  {journal} {\bibinfo  {journal} {JHEP}\ }\textbf {\bibinfo
  {volume} {02}},\ \bibinfo {pages} {212}},\ \Eprint
  {https://arxiv.org/abs/2206.14487} {arXiv:2206.14487 [hep-lat]} \BibitemShut
  {NoStop}%
\bibitem [{\citenamefont {Bonati}\ \emph
  {et~al.}(2021{\natexlab{a}})\citenamefont {Bonati}, \citenamefont {Franchi},
  \citenamefont {Pelissetto},\ and\ \citenamefont {Vicari}}]{Bonati:2021oqq}%
  \BibitemOpen
  \bibfield  {author} {\bibinfo {author} {\bibfnamefont {C.}~\bibnamefont
  {Bonati}}, \bibinfo {author} {\bibfnamefont {A.}~\bibnamefont {Franchi}},
  \bibinfo {author} {\bibfnamefont {A.}~\bibnamefont {Pelissetto}},\ and\
  \bibinfo {author} {\bibfnamefont {E.}~\bibnamefont {Vicari}},\ }\bibfield
  {title} {\bibinfo {title} {{Two-dimensional lattice SU(N$_{c}$) gauge
  theories with multiflavor adjoint scalar fields}},\ }\href
  {https://doi.org/10.1007/JHEP05(2021)018} {\bibfield  {journal} {\bibinfo
  {journal} {JHEP}\ }\textbf {\bibinfo {volume} {05}},\ \bibinfo {pages}
  {018}},\ \Eprint {https://arxiv.org/abs/2103.12708} {arXiv:2103.12708
  [hep-lat]} \BibitemShut {NoStop}%
\bibitem [{\citenamefont {Scammell}\ \emph {et~al.}(2020)\citenamefont
  {Scammell}, \citenamefont {Patekar}, \citenamefont {Scheurer},\ and\
  \citenamefont {Sachdev}}]{Scammell:2019erm}%
  \BibitemOpen
  \bibfield  {author} {\bibinfo {author} {\bibfnamefont {H.~D.}\ \bibnamefont
  {Scammell}}, \bibinfo {author} {\bibfnamefont {K.}~\bibnamefont {Patekar}},
  \bibinfo {author} {\bibfnamefont {M.~S.}\ \bibnamefont {Scheurer}},\ and\
  \bibinfo {author} {\bibfnamefont {S.}~\bibnamefont {Sachdev}},\ }\bibfield
  {title} {\bibinfo {title} {{Phases of SU(2) gauge theory with multiple
  adjoint Higgs fields in 2+1 dimensions}},\ }\href
  {https://doi.org/10.1103/PhysRevB.101.205124} {\bibfield  {journal} {\bibinfo
   {journal} {Phys. Rev. B}\ }\textbf {\bibinfo {volume} {101}},\ \bibinfo
  {pages} {205124} (\bibinfo {year} {2020})},\ \Eprint
  {https://arxiv.org/abs/1912.06108} {arXiv:1912.06108 [cond-mat.str-el]}
  \BibitemShut {NoStop}%
\bibitem [{\citenamefont {Bonati}\ \emph
  {et~al.}(2021{\natexlab{b}})\citenamefont {Bonati}, \citenamefont {Franchi},
  \citenamefont {Pelissetto},\ and\ \citenamefont {Vicari}}]{Bonati:2021tvg}%
  \BibitemOpen
  \bibfield  {author} {\bibinfo {author} {\bibfnamefont {C.}~\bibnamefont
  {Bonati}}, \bibinfo {author} {\bibfnamefont {A.}~\bibnamefont {Franchi}},
  \bibinfo {author} {\bibfnamefont {A.}~\bibnamefont {Pelissetto}},\ and\
  \bibinfo {author} {\bibfnamefont {E.}~\bibnamefont {Vicari}},\ }\bibfield
  {title} {\bibinfo {title} {{Three-dimensional lattice SU(Nc) gauge theories
  with multiflavor scalar fields in the adjoint representation}},\ }\href
  {https://doi.org/10.1103/PhysRevB.104.115166} {\bibfield  {journal} {\bibinfo
   {journal} {Phys. Rev. B}\ }\textbf {\bibinfo {volume} {104}},\ \bibinfo
  {pages} {115166} (\bibinfo {year} {2021}{\natexlab{b}})},\ \Eprint
  {https://arxiv.org/abs/2106.15152} {arXiv:2106.15152 [hep-lat]} \BibitemShut
  {NoStop}%
\bibitem [{\citenamefont {Proust}\ and\ \citenamefont
  {Taillefer}(2019)}]{Proust:2019eu}%
  \BibitemOpen
  \bibfield  {author} {\bibinfo {author} {\bibfnamefont {C.}~\bibnamefont
  {Proust}}\ and\ \bibinfo {author} {\bibfnamefont {L.}~\bibnamefont
  {Taillefer}},\ }\bibfield  {title} {\bibinfo {title} {The remarkable
  underlying ground states of cuprate superconductors},\ }\href
  {https://doi.org/10.1146/annurev-conmatphys-031218-013210} {\bibfield
  {journal} {\bibinfo  {journal} {Annual Review of Condensed Matter Physics}\
  }\textbf {\bibinfo {volume} {10}},\ \bibinfo {pages} {409} (\bibinfo {year}
  {2019})},\ \Eprint
  {https://arxiv.org/abs/https://doi.org/10.1146/annurev-conmatphys-031218-013210}
  {https://doi.org/10.1146/annurev-conmatphys-031218-013210} \BibitemShut
  {NoStop}%
\bibitem [{\citenamefont {Fujita}\ \emph {et~al.}(2014)\citenamefont {Fujita},
  \citenamefont {Kim}, \citenamefont {Lee}, \citenamefont {Lee}, \citenamefont
  {Hamidian}, \citenamefont {Firmo}, \citenamefont {Mukhopadhyay},
  \citenamefont {Eisaki}, \citenamefont {Uchida}, \citenamefont {Lawler},
  \citenamefont {Kim},\ and\ \citenamefont {Davis}}]{2014.Fujita}%
  \BibitemOpen
  \bibfield  {author} {\bibinfo {author} {\bibfnamefont {K.}~\bibnamefont
  {Fujita}}, \bibinfo {author} {\bibfnamefont {C.~K.}\ \bibnamefont {Kim}},
  \bibinfo {author} {\bibfnamefont {I.}~\bibnamefont {Lee}}, \bibinfo {author}
  {\bibfnamefont {J.}~\bibnamefont {Lee}}, \bibinfo {author} {\bibfnamefont
  {M.~H.}\ \bibnamefont {Hamidian}}, \bibinfo {author} {\bibfnamefont {I.~A.}\
  \bibnamefont {Firmo}}, \bibinfo {author} {\bibfnamefont {S.}~\bibnamefont
  {Mukhopadhyay}}, \bibinfo {author} {\bibfnamefont {H.}~\bibnamefont
  {Eisaki}}, \bibinfo {author} {\bibfnamefont {S.}~\bibnamefont {Uchida}},
  \bibinfo {author} {\bibfnamefont {M.~J.}\ \bibnamefont {Lawler}}, \bibinfo
  {author} {\bibfnamefont {E.-A.}\ \bibnamefont {Kim}},\ and\ \bibinfo {author}
  {\bibfnamefont {J.~C.}\ \bibnamefont {Davis}},\ }\bibfield  {title} {\bibinfo
  {title} {{Simultaneous Transitions in Cuprate Momentum-Space Topology and
  Electronic Symmetry Breaking}},\ }\href
  {https://doi.org/10.1126/science.1248783} {\bibfield  {journal} {\bibinfo
  {journal} {Science}\ }\textbf {\bibinfo {volume} {344}},\ \bibinfo {pages}
  {612} (\bibinfo {year} {2014})},\ \Eprint {https://arxiv.org/abs/1403.7788}
  {1403.7788} \BibitemShut {NoStop}%
\bibitem [{\citenamefont {He}\ \emph {et~al.}(2014)\citenamefont {He},
  \citenamefont {Yin}, \citenamefont {Zech}, \citenamefont {Soumyanarayanan},
  \citenamefont {Yee}, \citenamefont {Williams}, \citenamefont {Boyer},
  \citenamefont {Chatterjee}, \citenamefont {Wise}, \citenamefont {Zeljkovic},
  \citenamefont {Kondo}, \citenamefont {Takeuchi}, \citenamefont {Ikuta},
  \citenamefont {Mistark}, \citenamefont {Markiewicz}, \citenamefont {Bansil},
  \citenamefont {Sachdev}, \citenamefont {Hudson},\ and\ \citenamefont
  {Hoffman}}]{2014.He}%
  \BibitemOpen
  \bibfield  {author} {\bibinfo {author} {\bibfnamefont {Y.}~\bibnamefont
  {He}}, \bibinfo {author} {\bibfnamefont {Y.}~\bibnamefont {Yin}}, \bibinfo
  {author} {\bibfnamefont {M.}~\bibnamefont {Zech}}, \bibinfo {author}
  {\bibfnamefont {A.}~\bibnamefont {Soumyanarayanan}}, \bibinfo {author}
  {\bibfnamefont {M.~M.}\ \bibnamefont {Yee}}, \bibinfo {author} {\bibfnamefont
  {T.}~\bibnamefont {Williams}}, \bibinfo {author} {\bibfnamefont {M.~C.}\
  \bibnamefont {Boyer}}, \bibinfo {author} {\bibfnamefont {K.}~\bibnamefont
  {Chatterjee}}, \bibinfo {author} {\bibfnamefont {W.~D.}\ \bibnamefont
  {Wise}}, \bibinfo {author} {\bibfnamefont {I.}~\bibnamefont {Zeljkovic}},
  \bibinfo {author} {\bibfnamefont {T.}~\bibnamefont {Kondo}}, \bibinfo
  {author} {\bibfnamefont {T.}~\bibnamefont {Takeuchi}}, \bibinfo {author}
  {\bibfnamefont {H.}~\bibnamefont {Ikuta}}, \bibinfo {author} {\bibfnamefont
  {P.}~\bibnamefont {Mistark}}, \bibinfo {author} {\bibfnamefont {R.~S.}\
  \bibnamefont {Markiewicz}}, \bibinfo {author} {\bibfnamefont
  {A.}~\bibnamefont {Bansil}}, \bibinfo {author} {\bibfnamefont
  {S.}~\bibnamefont {Sachdev}}, \bibinfo {author} {\bibfnamefont {E.~W.}\
  \bibnamefont {Hudson}},\ and\ \bibinfo {author} {\bibfnamefont {J.~E.}\
  \bibnamefont {Hoffman}},\ }\bibfield  {title} {\bibinfo {title} {{Fermi
  Surface and Pseudogap Evolution in a Cuprate Superconductor}},\ }\href
  {https://doi.org/10.1126/science.1248221} {\bibfield  {journal} {\bibinfo
  {journal} {Science}\ }\textbf {\bibinfo {volume} {344}},\ \bibinfo {pages}
  {608} (\bibinfo {year} {2014})},\ \Eprint {https://arxiv.org/abs/1305.2778}
  {1305.2778} \BibitemShut {NoStop}%
\bibitem [{\citenamefont {Sachdev}\ \emph {et~al.}(2019)\citenamefont
  {Sachdev}, \citenamefont {Scammell}, \citenamefont {Scheurer},\ and\
  \citenamefont {Tarnopolsky}}]{Sachdev:2018nbk}%
  \BibitemOpen
  \bibfield  {author} {\bibinfo {author} {\bibfnamefont {S.}~\bibnamefont
  {Sachdev}}, \bibinfo {author} {\bibfnamefont {H.~D.}\ \bibnamefont
  {Scammell}}, \bibinfo {author} {\bibfnamefont {M.~S.}\ \bibnamefont
  {Scheurer}},\ and\ \bibinfo {author} {\bibfnamefont {G.}~\bibnamefont
  {Tarnopolsky}},\ }\bibfield  {title} {\bibinfo {title} {{Gauge theory for the
  cuprates near optimal doping}},\ }\href
  {https://doi.org/10.1103/PhysRevB.99.054516} {\bibfield  {journal} {\bibinfo
  {journal} {Phys. Rev. B}\ }\textbf {\bibinfo {volume} {99}},\ \bibinfo
  {pages} {054516} (\bibinfo {year} {2019})},\ \Eprint
  {https://arxiv.org/abs/1811.04930} {arXiv:1811.04930 [cond-mat.str-el]}
  \BibitemShut {NoStop}%
\bibitem [{\citenamefont {Zhang}\ \emph {et~al.}(2002)\citenamefont {Zhang},
  \citenamefont {Demler},\ and\ \citenamefont {Sachdev}}]{Zhang:2002zz}%
  \BibitemOpen
  \bibfield  {author} {\bibinfo {author} {\bibfnamefont {Y.}~\bibnamefont
  {Zhang}}, \bibinfo {author} {\bibfnamefont {E.}~\bibnamefont {Demler}},\ and\
  \bibinfo {author} {\bibfnamefont {S.}~\bibnamefont {Sachdev}},\ }\bibfield
  {title} {\bibinfo {title} {{Competing orders in a magnetic field: Spin and
  charge order in the cuprate superconductors}},\ }\href
  {https://doi.org/10.1103/PhysRevB.66.094501} {\bibfield  {journal} {\bibinfo
  {journal} {Phys. Rev. B}\ }\textbf {\bibinfo {volume} {66}},\ \bibinfo
  {pages} {094501} (\bibinfo {year} {2002})},\ \Eprint
  {https://arxiv.org/abs/cond-mat/0112343} {arXiv:cond-mat/0112343}
  \BibitemShut {NoStop}%
\bibitem [{\citenamefont {De~Prato}\ \emph {et~al.}(2006)\citenamefont
  {De~Prato}, \citenamefont {Pelissetto},\ and\ \citenamefont
  {Vicari}}]{DePrato:2006jx}%
  \BibitemOpen
  \bibfield  {author} {\bibinfo {author} {\bibfnamefont {M.}~\bibnamefont
  {De~Prato}}, \bibinfo {author} {\bibfnamefont {A.}~\bibnamefont
  {Pelissetto}},\ and\ \bibinfo {author} {\bibfnamefont {E.}~\bibnamefont
  {Vicari}},\ }\bibfield  {title} {\bibinfo {title} {{Spin-density-wave order
  in cuprates}},\ }\href {https://doi.org/10.1103/PhysRevB.74.144507}
  {\bibfield  {journal} {\bibinfo  {journal} {Phys. Rev. B}\ }\textbf {\bibinfo
  {volume} {74}},\ \bibinfo {pages} {144507} (\bibinfo {year} {2006})},\
  \Eprint {https://arxiv.org/abs/cond-mat/0601404} {arXiv:cond-mat/0601404}
  \BibitemShut {NoStop}%
\bibitem [{\citenamefont {Wegner}(1971)}]{Wegner:1971app}%
  \BibitemOpen
  \bibfield  {author} {\bibinfo {author} {\bibfnamefont {F.~J.}\ \bibnamefont
  {Wegner}},\ }\bibfield  {title} {\bibinfo {title} {{Duality in Generalized
  Ising Models and Phase Transitions Without Local Order Parameters}},\ }\href
  {https://doi.org/10.1063/1.1665530} {\bibfield  {journal} {\bibinfo
  {journal} {J. Math. Phys.}\ }\textbf {\bibinfo {volume} {12}},\ \bibinfo
  {pages} {2259} (\bibinfo {year} {1971})}\BibitemShut {NoStop}%
\bibitem [{\citenamefont {Sachdev}(2019)}]{Sachdev:2018ddg}%
  \BibitemOpen
  \bibfield  {author} {\bibinfo {author} {\bibfnamefont {S.}~\bibnamefont
  {Sachdev}},\ }\bibfield  {title} {\bibinfo {title} {{Topological order,
  emergent gauge fields, and Fermi surface reconstruction}},\ }\href
  {https://doi.org/10.1088/1361-6633/aae110} {\bibfield  {journal} {\bibinfo
  {journal} {Rept. Prog. Phys.}\ }\textbf {\bibinfo {volume} {82}},\ \bibinfo
  {pages} {014001} (\bibinfo {year} {2019})},\ \Eprint
  {https://arxiv.org/abs/1801.01125} {arXiv:1801.01125 [cond-mat.str-el]}
  \BibitemShut {NoStop}%
\bibitem [{sch()}]{scheurer_chi}%
  \BibitemOpen
  \href@noop {} {\bibinfo {title} {Private correspondence with {Mathias}
  {Scheurer}.}}\BibitemShut {Stop}%
\bibitem [{\citenamefont {Sarkar}\ \emph {et~al.}(2023)\citenamefont {Sarkar},
  \citenamefont {Catumba}, \citenamefont {Hiraguchi}, \citenamefont {Hou},
  \citenamefont {Jansen}, \citenamefont {Kao}, \citenamefont {Lin},\ and\
  \citenamefont {Ramos~Martinez}}]{Sarkar:2023lL}%
  \BibitemOpen
  \bibfield  {author} {\bibinfo {author} {\bibfnamefont {M.}~\bibnamefont
  {Sarkar}}, \bibinfo {author} {\bibfnamefont {G.}~\bibnamefont {Catumba}},
  \bibinfo {author} {\bibfnamefont {A.}~\bibnamefont {Hiraguchi}}, \bibinfo
  {author} {\bibfnamefont {G.~W.}\ \bibnamefont {Hou}}, \bibinfo {author}
  {\bibfnamefont {K.}~\bibnamefont {Jansen}}, \bibinfo {author} {\bibfnamefont
  {Y.-J.}\ \bibnamefont {Kao}}, \bibinfo {author} {\bibfnamefont {C.-J.~D.}\
  \bibnamefont {Lin}},\ and\ \bibinfo {author} {\bibfnamefont {A.}~\bibnamefont
  {Ramos~Martinez}},\ }\bibfield  {title} {\bibinfo {title} {{Study of SU(2)
  gauge theories with multiple Higgs fields in different representations}},\
  }\href {https://doi.org/10.22323/1.430.0388} {\bibfield  {journal} {\bibinfo
  {journal} {PoS}\ }\textbf {\bibinfo {volume} {LATTICE2022}},\ \bibinfo
  {pages} {388} (\bibinfo {year} {2023})}\BibitemShut {NoStop}%
\bibitem [{\citenamefont {Catumba}\ \emph {et~al.}(2024)\citenamefont
  {Catumba}, \citenamefont {Hiraguchi}, \citenamefont {Hou}, \citenamefont
  {Jansen}, \citenamefont {Kao}, \citenamefont {Lin}, \citenamefont {Ramos},\
  and\ \citenamefont {Sarkar}}]{Catumba:2023srt}%
  \BibitemOpen
  \bibfield  {author} {\bibinfo {author} {\bibfnamefont {G.}~\bibnamefont
  {Catumba}}, \bibinfo {author} {\bibfnamefont {A.}~\bibnamefont {Hiraguchi}},
  \bibinfo {author} {\bibfnamefont {G.~W.-S.}\ \bibnamefont {Hou}}, \bibinfo
  {author} {\bibfnamefont {K.}~\bibnamefont {Jansen}}, \bibinfo {author}
  {\bibfnamefont {Y.-J.}\ \bibnamefont {Kao}}, \bibinfo {author} {\bibfnamefont
  {C.-J.~D.}\ \bibnamefont {Lin}}, \bibinfo {author} {\bibfnamefont
  {A.}~\bibnamefont {Ramos}},\ and\ \bibinfo {author} {\bibfnamefont
  {M.}~\bibnamefont {Sarkar}},\ }\bibfield  {title} {\bibinfo {title} {{Study
  of 3-dimensional SU(2) gauge theory with adjoint Higgs as a model for cuprate
  superconductors}},\ }\href {https://doi.org/10.22323/1.453.0362} {\bibfield
  {journal} {\bibinfo  {journal} {PoS}\ }\textbf {\bibinfo {volume}
  {LATTICE2023}},\ \bibinfo {pages} {362} (\bibinfo {year} {2024})},\ \Eprint
  {https://arxiv.org/abs/2312.05537} {arXiv:2312.05537 [hep-lat]} \BibitemShut
  {NoStop}%
\end{thebibliography}%
 
\end{document}